\documentclass[12pt]{article}
\usepackage{amsmath,amssymb,amsthm,amsxtra,overpic,bbm,bm,epsfig,ulem}
\usepackage{color}
\usepackage{cite}
\usepackage{slashed}

\usepackage{hyperref}
\usepackage{url}

\def\thefootnote{\fnsymbol{footnote}}

\begin{document}

\vspace{0.2cm}

\begin{center}
{\large\bf Radiative decays of charged leptons as constraints of unitarity polygons \\
for active-sterile neutrino mixing and CP violation}
\end{center}

\vspace{0.2cm}

\begin{center}
{\bf Zhi-zhong Xing$^{1,2}$} and {\bf Di Zhang$^{1}$}
\footnote{E-mail: zhangdi@ihep.ac.cn (corresponding author)} \\
{\small $^{1}$Institute of High Energy Physics and School of Physical Sciences, \\
University of Chinese Academy of Sciences, Beijing 100049, China \\
$^{2}$Center of High Energy Physics, Peking University, Beijing 100871, China}
\end{center}

\vspace{2cm}
\begin{abstract}
We calculate the rates of radiative $\beta^- \to \alpha^- + \gamma$ decays for
$(\alpha, \beta) = (e, \mu)$, $(e, \tau)$ and $(\mu, \tau)$ by taking the {\it unitary}
gauge in the $(3+n)$ active-sterile neutrino mixing scheme, and make it clear that
constraints on the unitarity of the $3\times 3$ Pontecorvo-Maki-Nakagawa-Sakata (PMNS) matrix
$U$ extracted from $\beta^- \to \alpha^- + \gamma$ decays in the {\it minimal unitarity
violation} scheme differ from those obtained in the canonical seesaw mechanism with $n$ heavy
Majorana neutrinos by a factor $5/3$. In such a natural seesaw case we show that the rates
of $\beta^- \to \alpha^- + \gamma$ can be used to cleanly and strongly constrain the effective
apex of a unitarity polygon, and compare its geometry with the geometry of its three
sub-triangles formed by two vectors $U^{}_{\alpha i} U^*_{\beta i}$ and
$U^{}_{\alpha j} U^*_{\beta j}$ (for $i \neq j$) in the complex plane. We find that
the areas of such sub-triangles can be described in terms of the Jarlskog-like
invariants of CP violation ${\cal J}^{ij}_{\alpha\beta}$, and their small differences
signify slight unitarity violation of the PMNS matrix $U$.
\end{abstract}

\newpage

\def\thefootnote{\arabic{footnote}}
\setcounter{footnote}{0}
\setcounter{figure}{0}

\section{Introduction}

The discoveries of solar, atmospheric, reactor and accelerator neutrino
oscillations \cite{Zyla:2020zbs} have changed some of our previous ideas about
the standard model (SM) of particle physics, especially about its lepton
flavor sector. Now we are firmly convinced that neutrinos have mysteriously tiny masses and
lepton flavors are significantly mixed. Behind these two important observations
must be some kind of new physics beyond the SM \cite{Xing:2019vks}, which is responsible
for the origin of neutrino masses and may
have far-reaching implications for particle physics and cosmology.

In weak charged-current interactions it is the $3\times 3$
Pontecorvo-Maki-Nakagawa-Sakata (PMNS) matrix $U$ that
describes the effects of lepton flavor mixing and CP violation \cite{Pontecorvo:1957cp,Maki:1962mu,Pontecorvo:1967fh}. The smallest element
of $U$ is found to be $|U^{}_{e3}| \simeq 0.15$ \cite{Adey:2018zwh}, much larger
than $|V^{}_{ub}| \simeq 3.6 \times 10^{-3}$ --- the smallest element of the
$3\times 3$ Cabibbo-Kobayashi-Maskawa (CKM) quark flavor mixing matrix $V$
\cite{Cabibbo:1963yz,Kobayashi:1973fv}. Moreover, the PMNS matrix $U$ is very
likely to accommodate significant CP violation in the lepton sector, as indicated
by the latest T2K measurement at almost the $3\sigma$ confidence level \cite{Abe:2019vii}.
Given the unitarity of $U$, there are three Dirac-type unitarity triangles
defined by the orthogonality conditions \cite{Fritzsch:1999ee}
\footnote{There are also three Majorana-type triangles
defined by the orthogonality relations
$U^{}_{e i} U^*_{e j} + U^{}_{\mu i} U^*_{\mu j} + U^{}_{\tau i} U^*_{\tau j} = 0$
in the complex plane \cite{Fritzsch:1999ee,AguilarSaavedra:2000vr}, where
$(i, j) = (1, 2)$, $(2, 3)$ or $(3, 1)$. The configuration of each triangle
is sensitive to the relative phases of three Majorana neutrino mass eigenstates $\nu^{}_i$ (for
$i = 1, 2, 3$) \cite{Xing:2015wzz}, but it is irrelevant to the radiative decays of
charged leptons  and hence will not be discussed in this paper.}
\begin{eqnarray}
U^{}_{\alpha 1} U^*_{\beta 1} + U^{}_{\alpha 2} U^*_{\beta 2} + U^{}_{\alpha 3} U^*_{\beta 3} = 0 \;
\end{eqnarray}
in the complex plane, where $(\alpha, \beta) = (e, \mu)$, $(\mu, \tau)$ or $(\tau, e)$.
The geometry of each triangle depends only upon the Dirac phase $\delta$ of $U$ which
gives rise to leptonic CP violation in those ``appearance"-type long-baseline
neutrino oscillations as recently constrained by the T2K experiment \cite{Abe:2019vii}. So far
quite a lot of attention has been paid to leptonic unitarity triangles in vacuum or
in matter \cite{Xing:2005gk,Luo:2011mm,Xing:2015wzz,Esteban:2016qun,Zhu:2018dvj,
Xing:2019tsn,Ellis:2020ehi,Ellis:2020hus},
partly because such a geometric language has proved to be very
successful in the quark sector to intuitively describe the most salient features of flavor mixing
and CP violation.

\begin{figure}[t]
\centering
\includegraphics[width=18cm]{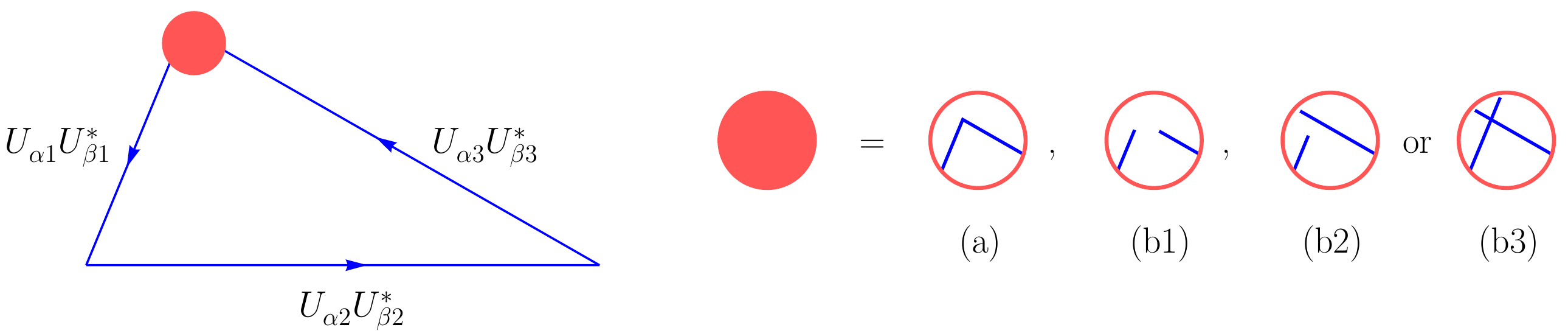}
\caption{The schematic plot of an {\it effective} Dirac-type unitarity triangle
and four typical topologies of its apex in the complex plane based on the $(3+n)$
active-sterile neutrino mixing scheme, where $(\alpha, \beta) = (e, \mu)$,
$(\mu, \tau)$ or $(\tau, e)$. As for the {\it effective} apex in red, pattern (a)
corresponds to the exact unitarity triangle of three active neutrinos
defined by Eq.~(1); and pattern (b1), (b2) or (b3) stands for a unitarity polygon
of $(3+n)$ neutrinos defined by Eq.~(3).}
\label{triangle}
\end{figure}

But unlike the CKM matrix $V$, whose unitarity is guaranteed by the SM itself, whether
the PMNS matrix $U$ is exactly unitary or not depends on the origin of neutrino masses.
From a theoretical point of view, the most natural and popular mechanism of generating
finite but tiny neutrino masses is the canonical seesaw mechanism
\cite{Minkowski:1977sc,Yanagida:1979as,GellMann:1980vs,Glashow:1979nm,Mohapatra:1979ia}
--- an extension of the SM by adding $n$ heavy (right-handed) neutrino fields
and allowing lepton number violation. Such $\rm SU(2)^{}_{\rm L}$-singlet neutrino
fields are coupled with the leptonic $\rm SU(2)^{}_{\rm L}$ doublet and the Higss
doublet as an SM-like Yukawa interaction term, and they may also form a gauge-invariant
but lepton-number-violating Majorana mass term with their own charge-conjugated
counterparts. After spontaneous gauge symmetry
breaking, the resultant neutrino mass matrix is a $\left(3 + n\right) \times
\left(3 + n\right)$ symmetric matrix whose eigenvectors correspond to three light
Majorana neutrino fields $\nu^{}_i$ (for $i = 1, 2, 3$) and $n$ heavy Majorana neutrino
fields $N^{}_j$ (for $j = 1, 2, \cdots, n$). In this case the mixing between
light (active) and heavy (sterile) neutrino flavors is described by a $3\times n$
matrix $R$ \cite{Xing:2007zj,Xing:2011ur}, and it is correlated with the $3\times 3$ PMNS matrix
$U$ via $U U^\dagger + R R^\dagger = I$. Since both $U$ and $R$ are the sub-matrices
of a $\left(3 + n\right) \times \left(3 + n\right)$ unitary matrix $\cal U$ used to diagonalize
the overall $\left(3 + n\right) \times \left(3 + n\right)$ neutrino mass matrix,
neither of them is exactly unitary. The standard weak charged-current interactions of
three charged leptons and $(3+n)$ neutrinos turn out to be \cite{Xing:2007zj,Xing:2011ur}
\begin{eqnarray}
{\cal L}^{}_{\rm cc} = \frac{g}{\sqrt{2}} \ \overline{\left(e \quad
\mu \quad \tau\right)^{}_{\rm L}} \ \gamma^\mu \left[ U \left(
\begin{matrix} \nu^{}_1 \cr \nu^{}_2 \cr \nu^{}_3 \end{matrix}
\right)^{}_{\hspace{-0.1cm}\rm L} + R \left( \begin{matrix} N^{}_1 \cr \vdots \cr
N^{}_n \end{matrix} \right)^{}_{\hspace{-0.1cm}\rm L} \right] W^-_\mu + {\rm h.c.} \; ,
\end{eqnarray}
where both the charged leptons and neutrinos are in their mass eigenstates, and the correlation
between $R$ and $U$ has been given above (i.e., $UU^\dagger = I - RR^\dagger$).
As a straightforward consequence, the three Dirac-type unitarity triangles defined in Eq.~(1)
are now replaced with three unitarity polygons defined by the orthogonality relations
\begin{eqnarray}
U^{}_{\alpha 1} U^*_{\beta 1} + U^{}_{\alpha 2} U^*_{\beta 2} + U^{}_{\alpha 3} U^*_{\beta 3}
= - \sum^n_{i = 1} R^{}_{\alpha i} R^*_{\beta i} \;
\end{eqnarray}
in the complex plane, where $(\alpha, \beta) = (e, \mu)$, $(\mu, \tau)$ or $(\tau, e)$.
In other words, the unitarity of $U$ is violated due to $R \neq 0$. Since the strength of
active-sterile flavor mixing is expected to be very small in a natural seesaw model,
the departure of a unitarity polygon from the corresponding unitarity triangle should also be
very small and can serve as a clear signal of new physics if it is finally measured
at low energies. Figure~\ref{triangle} schematically illustrates an {\it effective} Dirac-type
unitarity triangle and four typical topologies of its apex in the complex plane, where pattern (a)
corresponds to exact unitarity in the standard case and pattern (b1), (b2) or (b3) stands
for slight unitarity violation in the presence of active-sterile neutrino mixing.
Then the question becomes which weak-interaction process at low energies is appropriate for
constraining a given unitarity polygon and probing its deviation
from the corresponding unitarity triangle.

In the present paper we are going to show that the radiative decays of charged
leptons, denoted as $\beta^- \to \alpha^- + \gamma$ for $(\alpha, \beta) = (e, \mu)$,
$(e, \tau)$ or $(\mu, \tau)$, can be used to impose a straightforward and strong
constraint on the effective apex of a given unitarity polygon in the canonical seesaw
mechanism with the heavy Majorana neutrino masses $M^{}_i$ (for $i = 1, 2, \cdots, n$)
being far above the charged vector boson mass $M^{}_W$. It is worth pointing out that
our work is different from those previous studies (see Ref.~\cite{Lindner:2016bgg}
for a recent review with extensive references) in the following three aspects.
\begin{itemize}
\item     We calculate the rates of lepton-flavor-violating
$\beta^- \to \alpha^- + \gamma$ decays mediated by both three active neutrinos and $n$
sterile neutrinos via Eq.~(2) in the {\it unitary} gauge, where the sterile neutrinos can
in general be either heavy or light (i.e., one is unnecessarily subject to the canonical
seesaw mechanism). Our result is consistent with those obtained in Refs.~\cite{Ilakovac:1994kj,Alonso:2012ji}
\footnote{The {\it Feynman-'t Hooft} gauge has been adopted in Ref.~\cite{Ilakovac:1994kj}
for the calculations of radiative $\beta^- \to \alpha^- + \gamma$ decays, but
In Ref.~\cite{Alonso:2012ji} the authors made no mention of which gauge has been used to
carry out the Feynman-diagram calculations. Here we make use of the {\it unitary} gauge
instead of other gauges, so as to minimize the number of Feynman diagrams
and make an independent crosscheck of the results obtained in the literature.}.
Switching off the sterile neutrinos, we may immediately reproduce
the pioneering results achieved in 1977 \cite{Minkowski:1977sc,Petcov:1976ff,Bilenky:1977du,
Cheng:1976uq,Marciano:1977wx,Lee:1977qz,Lee:1977tib}. We make it clear that constraints on
the unitarity of $U$ extracted from $\beta^- \to \alpha^- + \gamma$ decays in the
so-called {\it minimal unitarity violation} scheme (see, e.g., Refs.~\cite{Antusch:2006vwa,Antusch:2014woa,Calibbi:2017uvl})
are different from those obtained in the canonical seesaw mechanism with $n$ heavy Majorana
neutrinos by a factor $5/3$, simply because there is a constant which dominates the
loop function in the former case but it is cancelled out in the latter case due to the
unitarity condition $UU^\dagger + RR^\dagger = I$.

\item     We illustrate how the loop function $G^{}_\gamma (x^{}_i)$ evolves with
$x^{}_i \equiv \lambda^2_i/M^{2}_W$, where $\lambda^{}_i$ represents an arbitrary neutrino
mass (for $i = 1, 2, \cdots, n+3$), and identify the two asymptotic convergence limits
$G^{}_\gamma (x^{}_i) \to -1/3$ for $x^{}_i \gg 1$ and $G^{}_\gamma (x^{}_i) \to -5/6$
for $x^{}_i \ll 1$. In the former case we demonstrate that the rates of
$\beta^- \to \alpha^- + \gamma$ can be used to cleanly constrain the effective apex
of a unitarity polygon shown by Figure~\ref{triangle} in the canonical seesaw mechanism,
and in the latter case we propose to define three new effective neutrino masses
which are directly sensitive to charged lepton flavor violation. Some numerical
results are also obtained in these two cases.

\item     We explore the geometry of a given unitarity polygon as compared with the
geometry of its three sub-triangles $\triangle^{ij}_{\alpha\beta}$ formed by
two vectors $U^{}_{\alpha i} U^*_{\beta i}$ and $U^{}_{\alpha j} U^*_{\beta j}$
(for $i \neq j$) in the complex plane. The areas of such sub-triangles can be
described in terms of the Jarlskog-like invariants of CP violation \cite{Jarlskog:1985ht}
\begin{eqnarray}
{\cal J}^{ij}_{\alpha\beta} = {\rm Im}\left(U^{}_{\alpha i} U^{}_{\beta j}
U^*_{\alpha j} U^*_{\beta i}\right) \; ,
\end{eqnarray}
where $(\alpha, \beta) = (e, \mu)$, $(\mu, \tau)$ or $(\tau,e)$ and
$(i, j) = (1, 2)$, $(2, 3)$ or $(3, 1)$, and their differences signify slight
unitarity violation of the $3\times 3$ PMNS matrix $U$.
\end{itemize}
Therefore, we expect that the main results of our study will be useful for testing
unitarity of the PMNS matrix and probing possible active-sterile neutrino mixing
in the era of precision measurements of neutrino oscillations, charged lepton
flavor violation and even lepton number violation.

The remaining parts of this paper are organized as follows. In section 2 we calculate
the rates of radiative $\beta^- \to \alpha^- + \gamma$ decays in the $(3+n)$ active-sterile
neutrino mixing scheme by making use of the {\it unitary} gauge, and discuss some salient features
of the loop function. Section 3 is devoted to some straightforward but strong constraints
on a given unitarity polygons in a natural seesaw framework, and to some explicit
discussions about its sub-triangles $\triangle^{ij}_{\alpha\beta}$ by taking the most
popular $(3+3)$ active-sterile neutrino mixing scenario for example.
We summarize our main results in section 4.

\section{Radiative $\beta^- \to \alpha^- + \gamma$ decays}

\begin{figure}[t!]
	\centering
	\includegraphics[width=\linewidth]{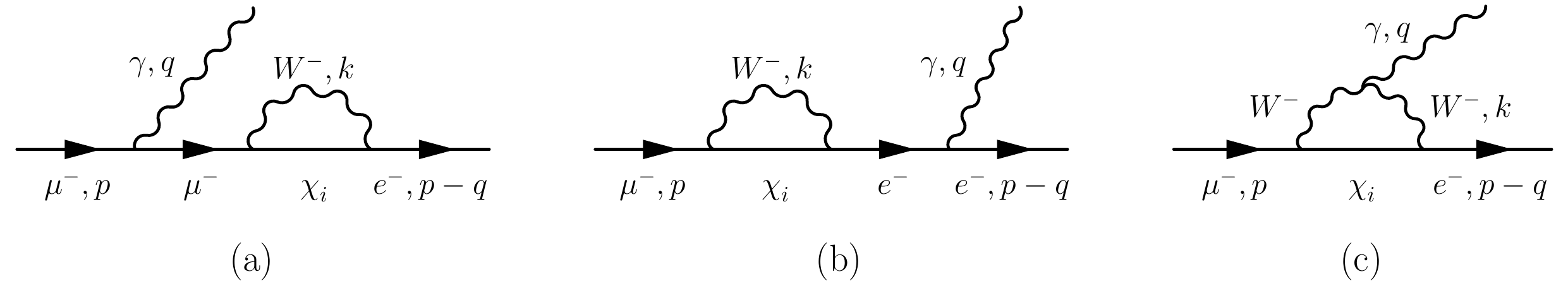}
	\caption{The one-loop Feynman diagrams contributing to $\mu^- \to e^- + \gamma$ in the
		{\it unitary} gauge, where $\chi^{}_i$ can be either three active neutrinos (i.e.,
		$\chi^{}_i = \nu^{}_i$ with mass $m^{}_i$ for $i = 1, 2, 3$) or $n$ sterile neutrinos
		(i.e., $\chi^{}_{i+3} = N^{}_i$ with mass $M^{}_i$ for $i = 1, 2, \cdots, n$)
		in the $(3+n)$ active-sterile flavor mixing scheme.}
	\label{meg}
\end{figure}

Let us take $\mu^- \to e^- + \gamma$ as an example to show how to calculate the
rates of radiative $\beta^- \to \alpha^- + \gamma$ decays for
$(\alpha, \beta) = (e, \mu)$, $(e, \tau)$ or $(\mu, \tau)$. Given the $(3+n)$
active-sterile neutrino mixing scheme and the weak charged-current interactions of such
neutrinos as described by Eq.~(2), the lowest-order (one-loop) Feynman diagrams
which contribute to $\mu^- \to e^- + \gamma$ in the {\it unitary} gauge are plotted in
Figure~\ref{meg}, where $\chi^{}_i$ with mass $\lambda^{}_i$ (for $i=1, 2, \cdots, 3+n$)
represents an arbitrary neutrino field under discussion no matter whether it is active or sterile
and whether it is light or heavy. For the sake of simplicity,
one may tentatively use ${\cal U}^{}_{\alpha i}$ (for $i = 1, 2, \cdots, 3+n$) to universally
describe $U^{}_{\alpha i}$ (for $i = 1, 2, 3$) and $R^{}_{\alpha i}$
(for $i = 1, 2, \cdots, n$), since $U$ and $R$ are respectively the upper-left $3\times 3$
and upper-right $3\times n$ sub-matrices of the $(3+n) \times (3+n)$ unitary
matrix $\cal U$. With the help of the notations $\chi^{}_i$, $\lambda^{}_i$ and
${\cal U}^{}_{\alpha i}$, the active and sterile neutrinos can be treated on the
same footing in calculating the rate of $\mu^- \to e^- + \gamma$.

In the {\it unitary} gauge and $D$ dimensions ($D \equiv 4-\varepsilon$), the three decay
amplitudes for $\mu^- \to e^- + \gamma$ shown in Figure~\ref{meg} can be expressed as
\begin{eqnarray}
{\rm i} \mathcal{M}^{}_{\rm a} \hspace{-0.2cm}&=&\hspace{-0.2cm} - \frac{1}{2} e g^2
\epsilon^\ast_\rho \left( q \right) \sum^{3+n}_{i=1} \mathcal{U}^{}_{ei} \mathcal{U}^\ast_{\mu i}~
\mu^{ \frac{3 }{2} \varepsilon } \int \frac{{\rm d}^D k}{\left(2\pi\right)^D}
\overline{u}\left( p-q \right) \gamma^\mu P^{}_{\rm L} \frac{ \slashed{p} -\slashed{q}
- \slashed{k} + \lambda^{}_i }{\left( p-q-k \right)^2 - \lambda^2_i} \gamma^\nu P^{}_{\rm L}
\nonumber \\
\hspace{-0.2cm}&&\hspace{-0.2cm} \times \frac{ \slashed{p}-\slashed{q} + m^{}_\mu }
{\left(p-q\right)^2 -m^2_\mu} \gamma^\rho u \left( p \right) \left( g^{}_{\mu\nu}
- \frac{k^{}_\mu k^{}_\nu}{M^2_W} \right) \frac{1}{k^2-M^2_W} \;,
\nonumber \\
{\rm i} \mathcal{M}^{}_{\rm b} \hspace{-0.2cm}&=&\hspace{-0.2cm} - \frac{1}{2} e g^2
\epsilon^\ast_\rho \left( q \right) \sum^{3+n}_{i=1} \mathcal{U}^{}_{ei} \mathcal{U}^\ast_{\mu i}~
\mu^{ \frac{3 }{2} \varepsilon } \int \frac{{\rm d}^D k}{\left(2\pi\right)^D}
\overline{u}\left( p-q \right) \gamma^\rho \frac{ \slashed{p} + m^{}_e }{p^2
- m^2_e} \gamma^\mu P^{}_{\rm L}
\nonumber \\
\hspace{-0.2cm}&&\hspace{-0.2cm} \times \frac{ \slashed{p} - \slashed{k} +\lambda^{}_i }
{\left( p-k \right)^2 - \lambda^2_i} \gamma^\nu P^{}_{\rm L} u \left( p \right) \left(
g^{}_{\mu\nu} - \frac{k^{}_\mu k^{}_\nu}{M^2_W} \right) \frac{1}{k^2-M^2_W} \;,
\nonumber \\
{\rm i} \mathcal{M}^{}_{\rm c} \hspace{-0.2cm}&=&\hspace{-0.2cm}  \frac{1}{2} e g^2
\epsilon^\ast_\rho \left( q \right) \sum^{3+n}_{i=1} \mathcal{U}^{}_{ei} \mathcal{U}^\ast_{\mu i}~
\mu^{ \frac{3 }{2} \varepsilon } \int \frac{{\rm d}^D k}{\left(2\pi\right)^D}
\overline{u}\left( p-q \right) \gamma^\mu P^{}_{\rm L} \frac{ \slashed{p} -\slashed{q}
-\slashed{k} + \lambda^{}_i }{\left( p-q-k \right)^2 -\lambda^2_i} \gamma^\nu P^{}_{\rm L}
u \left(p\right)
\nonumber \\
\hspace{-0.2cm}&&\hspace{-0.2cm}  \times \left( g^{}_{\mu\sigma} - \frac{k^{}_\mu
k^{}_\sigma}{M^2_W} \right) \frac{1}{k^2-M^2_W} \left[ - g^{\sigma\lambda} \left(
q+2k \right)^\rho + g^{\lambda\rho} \left( 2q+k \right)^\sigma + g^{\rho\sigma}
\left( k-q \right)^\lambda \right]
\nonumber \\
\hspace{-0.2cm}&&\hspace{-0.2cm} \times \left[ g^{}_{\lambda \nu} - \frac{\left(
q+k \right)^{}_\lambda \left( q+k \right)^{}_\nu}{M^2_W} \right] \frac{1}{\left(
q+k \right)^2 - M^2_W} \;,
\end{eqnarray}
where $\mu$ is an arbitrary mass-dimension-one parameter to keep the electromagnetic and
weak coupling constants $e$ and $g$ to be dimensionless in $D$ dimensions.
A more specific introduction of $\mu$, together with the algebra and traces of the Dirac
matrices in $D$ dimensions, has been given in appendix A. In appendix B
we have explicitly figured out the integrals in Eq.~(5) with the help of
the on-shell conditions $p^2 = m^2_\mu$, $q^2 = 0$ and $p \cdot
q = \left(m^2_\mu - m^2_e \right)/2$, the physical polarizations
for the external photon $\epsilon \left(q\right) \cdot q = 0$ and
the well-known Passarino-Veltman integrals. Then we take the limit $D \to 4$ (i.e.,
$\varepsilon \to 0$) and arrive at the results
\begin{eqnarray}
{\rm i} \mathcal{M}^{}_{\rm a} \hspace{-0.2cm}&=&\hspace{-0.2cm} \frac{{\rm i} e g^2}{2\left( 4\pi
	\right)^2}  \epsilon^\ast_\rho \left( q \right) \overline{u} \left(p-q\right) \frac{m^{}_e
	\left( m^{}_e P^{}_{\rm R} + m^{}_\mu P^{}_{\rm L} \right)}{m^2_\mu - m^2_e} \gamma^\rho u \left(p\right)
\sum^{3+n}_{i=1} \mathcal{U}^{}_{ei} \mathcal{U}^\ast_{\mu i}~ \mu^{ \frac{1}{2} \varepsilon}
\nonumber \\
\hspace{-0.2cm}&&\hspace{-0.2cm} \times \left\{ \frac{1}{2} \left( 3x^{}_i - \frac{m^2_e}{M^2_W}
\right) \left( \Delta - \ln \frac{M^2_W}{\mu^2} \right) + \frac{5x^2_i - 5x^{}_i -6}
{4\left(x^{}_i-1\right)} -\frac{3x^2_i \left(x^{}_i-2\right)}{2\left(x^{}_i-1\right)^2}
\ln x^{}_i \right.
\nonumber
\\
\hspace{-0.2cm}&&\hspace{-0.2cm} - \left. \frac{m^2_e}{M^2_W} \left[ \frac{x^3_i - 3x^2_i +
	45x^{}_i - 7}{ 12\left( x^{}_i-1 \right)^3} - \frac{x^2_i \left( x^2_i - 4x^{}_i + 9 \right)}
{2\left( x^{}_i-1 \right)^4} \ln x^{}_i \right] \right\} \;,
\end{eqnarray}
and
\begin{eqnarray}
{\rm i} \mathcal{M}^{}_{\rm b} \hspace{-0.2cm}&=&\hspace{-0.2cm} - \frac{{\rm i} e g^2}{2\left( 4\pi
\right)^2}  \epsilon^\ast_\rho \left( q \right) \overline{u} \left(p-q\right) \frac{m^{}_\mu
\left( m^{}_\mu P^{}_{\rm R} + m^{}_e P^{}_{\rm L} \right)}{m^2_\mu - m^2_e} \gamma^\rho u \left(p\right)
\sum^{3+n}_{i=1} \mathcal{U}^{}_{ei} \mathcal{U}^\ast_{\mu i}~ \mu^{ \frac{1}{2} \varepsilon}
\nonumber \\
\hspace{-0.2cm}&&\hspace{-0.2cm} \times \left\{ \frac{1}{2} \left( 3x^{}_i - \frac{m^2_\mu}{M^2_W}
\right) \left( \Delta - \ln \frac{M^2_W}{\mu^2} \right) + \frac{5x^2_i - 5x^{}_i -6}{4\left(x^{}_i-1\right)}
-\frac{3x^2_i \left(x^{}_i-2\right)}{2\left(x^{}_i-1\right)^2} \ln x^{}_i \right.
\nonumber
\\
\hspace{-0.2cm}&&\hspace{-0.2cm} - \left. \frac{m^2_\mu}{M^2_W} \left[ \frac{x^3_i - 3x^2_i + 45x^{}_i - 7}
{ 12\left( x^{}_i-1 \right)^3} - \frac{x^2_i \left( x^2_i - 4x^{}_i + 9 \right)}{2\left( x^{}_i-1 \right)^4}
\ln x^{}_i \right] \right\} \;,
\end{eqnarray}
as well as
\begin{eqnarray}
{\rm i} \mathcal{M}^{}_{\rm c} \hspace{-0.2cm}&=&\hspace{-0.2cm} - \frac{{\rm i} e g^2}{2\left( 4\pi
\right)^2 M^2_W} \epsilon^\ast_\rho \left( q \right) \sum^{3+n}_{i=1} \mathcal{U}^{}_{ei}
\mathcal{U}^\ast_{\mu i}~ \mu^{ \frac{1}{2} \varepsilon} \left\{\phantom{\frac{1}{1}} \hspace{-0.3cm}
G^{}_\gamma \left( x^{}_i \right) \overline{u} \left(p-q\right) {\rm i} \sigma^{\rho\lambda} q^{}_\lambda
\left( m^{}_e P^{}_{\rm L} + m^{}_\mu P^{}_{\rm R} \right) u \left(p\right) \right.
\nonumber \\
\hspace{-0.2cm}&&\hspace{-0.2cm}  + \overline{u} \left(p-q\right) \left[ m^{}_e m^{}_\mu  P^{}_{\rm L}
+ \left( m^2_e + m^2_\mu \right) P^{}_{\rm R} \right] \gamma^\rho u \left(p\right) \left[ \frac{x^3_i
- 3x^2_i + 45x^{}_i - 7 }{12\left( x^{}_i - 1 \right)^3} \right.
\nonumber \\
\hspace{-0.2cm}&&\hspace{-0.2cm} - \left. \frac{x^2_i \left( x^2_i - 4x^{}_i + 9 \right)}{2\left( x^{}_i
- 1 \right)^4} \ln x^{}_i \right] - \overline{u} \left(p-q\right) P^{}_{\rm R} \gamma^\rho u \left(p\right)
M^2_{W} \left[ \frac{5x^2_i - 5x^{}_i - 6}{4\left( x^{}_i - 1 \right)} \right.
\nonumber \\
\hspace{-0.2cm}&&\hspace{-0.2cm} - \left.\left. \frac{3x^2_i \left( x^{}_i - 2 \right)}{2\left( x^{}_i
- 1 \right)^2} \ln x^{}_i \right] + \frac{1}{2} \overline{u} \left(p-q\right) \left[ m^{}_e m^{}_\mu
P^{}_{\rm L} + \left( m^2_e + m^2_\mu - 3x^{}_i M^2_W \right) P^{}_{\rm R} \right] \gamma^\rho u
\left(p\right) \right.
\nonumber \\
\hspace{-0.2cm}&&\hspace{-0.2cm} \times \left. \left( \Delta - \ln \frac{M^2_W}{\mu^2} \right) \right\}
\end{eqnarray}
where $x^{}_i \equiv \lambda^2_i / M^2_W$,
$\Delta \equiv 2/\varepsilon - \gamma^{}_{\rm E} + \ln \left(4\pi\right)$ with
$\gamma^{}_{\rm E}$ being Euler's constant, and
\begin{eqnarray}
G^{}_\gamma \left(x^{}_i\right) = - \frac{5}{6} -
\frac{2x^3_i + 5x^2_i - x^{}_i}{4\left(1-x^{}_i\right)^3} -
\frac{3x^3_i}{2\left(1-x^{}_i\right)^4} \ln x^{}_i \;.
\end{eqnarray}
Note that both ${\rm i} \mathcal{M}^{}_{a} + {\rm i} \mathcal{M}^{}_{b}$ and
${\rm i} \mathcal{M}^{}_{c}$ keep unchanged under the exchange
$m^{}_e \leftrightarrow m^{}_\mu$, and the former can be explicitly expressed as
\begin{eqnarray}
{\rm i} \mathcal{M}_a + {\rm i} \mathcal{M}_b \hspace{-0.2cm}&=&\hspace{-0.2cm}
\frac{{\rm i} e g^2}{2\left( 4\pi \right)^2 M^2_W} \epsilon^\ast_\rho \left( q \right)
\sum\limits^{3+n}_{i=1} \mathcal{U}^{}_{ei} \mathcal{U}^\ast_{\mu i} ~
\mu^{ \frac{1}{2} \varepsilon} \left\{ \phantom{\frac{1}{1}}\hspace{-0.3cm}
\overline{u} \left(p-q\right) \left[ m^{}_e m^{}_\mu  P^{}_{\rm L} +
\left( m^2_e + m^2_\mu \right) P^{}_{\rm R} \right] \gamma^\rho u \left(p\right) \right.
\nonumber \\
\hspace{-0.2cm}&&\hspace{-0.2cm} \times \left[ \frac{x^3_i - 3x^2_i + 45x^{}_i - 7 }
{12\left( x^{}_i - 1 \right)^3} - \frac{x^2_i \left( x^2_i - 4x^{}_i + 9 \right)}
{2\left( x^{}_i - 1 \right)^4} \ln x^{}_i \right] - \overline{u} \left(p-q\right)
P^{}_{\rm R} \gamma^\rho u \left(p\right) M^2_{W}
\nonumber \\
\hspace{-0.2cm}&&\hspace{-0.2cm} \times \left[ \frac{5x^2_i - 5x^{}_i - 6}{4\left(
x^{}_i - 1 \right)} - \frac{3x^2_i \left( x^{}_i - 2 \right)}{2\left( x^{}_i - 1
\right)^2} \ln x^{}_i \right] + \frac{1}{2} \overline{u} \left(p-q\right) \left[
m^{}_e m^{}_\mu  P^{}_{\rm L} \right.
\nonumber \\
\hspace{-0.2cm}&&\hspace{-0.2cm} + \left.\left. \left( m^2_e + m^2_\mu - 3x^{}_i M^2_W
\right) P^{}_{\rm R} \right] \gamma^\rho u \left(p\right) \left( \Delta -
\ln \frac{M^2_W}{\mu^2} \right) \right\} \;.
\end{eqnarray}
It becomes clear that the terms of ${\rm i} \mathcal{M}^{}_{a} + {\rm i} \mathcal{M}^{}_{b}$ in
Eq.~(10) can exactly eliminate the terms of ${\rm i} \mathcal{M}^{}_{c}$ shown in the last
four rows of Eq.~(8), including the divergent terms and $\mu$-dependent terms. This observation
means that it is {\it unnecessary} to invoke any specific renormalization scheme to assure the
total decay amplitude $\mathcal{M} = \mathcal{M}^{}_{\rm a} + \mathcal{M}^{}_{\rm b} +
\mathcal{M}^{}_{\rm c}$ to be finite, and thus the overall factor $\mu^{\varepsilon/2}$ can be
simply removed from $\mathcal{M}$. As a result,
\begin{eqnarray}
{\rm i} \mathcal{M} \hspace{-0.2cm}&=&\hspace{-0.2cm} \frac{-{\rm i} e g^2}{2\left( 4\pi \right)^2 M^2_W}
\sum\limits^{3+n}_{i=1} \mathcal{U}^{}_{ei} \mathcal{U}^\ast_{\mu i} G^{}_\gamma \left(x^{}_i\right)
\left[\phantom{\frac{1}{1}} \hspace{-0.25cm}
\epsilon^\ast_\rho \left(q\right) \overline{u} \left(p-q\right) {\rm i} \sigma^{\rho\lambda}
q^{}_\lambda \left( m^{}_e P^{}_{\rm L} + m^{}_\mu P^{}_{\rm R} \right) u
\left(p\right) \phantom{\frac{1}{1}} \hspace{-0.3cm} \right] \; .
\end{eqnarray}
We conclude that our results in Eqs.~(9) and (11) are consistent with those obtained in
Refs.~\cite{Cheng:1980tp,Inami:1980fz,Lim:1981kv,Langacker:1988up,Ilakovac:1994kj,
Alonso:2012ji,Lindner:2016bgg}, but some necessary comments and clarifications are in order.
\begin{itemize}
\item      In Ref.~\cite{Cheng:1980tp} a similar loop function $F\left( x^{}_i \right)$
for $\mu^- \to e^- + \gamma$ has been obtained in the canonical seesaw framework,
but it is presented in the form of several integrals. After explicitly figuring out
those integrals, we arrive at
\begin{eqnarray}
F \left(x^{}_i\right) = \frac{10}{3} + \frac{2x^3_i + 5x^2_i - x^{}_i}
{\left(1-x^{}_i\right)^3} + \frac{6x^3_i}{\left(1-x^{}_i\right)^4} \ln x^{}_i \; .
\end{eqnarray}
It is clear that $G^{}_\gamma\left(x^{}_i\right) = -F\left(x^{}_i\right)/4$ holds, and
the factor $-1/4$ can be compensated by an additional factor $-4$ in the total decay amplitude
$\mathcal{M}$ of ours. Therefore, our result is fully in agreement with the one obtained in
Ref.~\cite{Cheng:1980tp}.

\item      Note that the constant in either $G^{}_\gamma \left(x^{}_i\right)$ or
$F \left(x^{}_i\right)$ can be exactly eliminated due to the unitarity of $\mathcal{U}$
in the $(3+n)$ active-sterile neutrino mixing scheme.
That is why a result which is equivalent to $-\left[F\left(x^{}_i\right) -10/3\right]$
has been given in Ref.~\cite{Langacker:1988up}. If one is subject to the {\it minimal
unitarity violation} scheme as considered in Refs.~\cite{Antusch:2006vwa,Antusch:2014woa}, the
calculations and results shown in Eqs.~(5)---(11) remain valid after $n = 0$ and ${\cal U} = U$
are taken; but one should keep in mind that in this case ${\cal U} = U$ is not unitary due to the
influence of one or more dimension-6 operators, and the corresponding loop function is actually
$F\left(x^{}_i\right)$ in Eq.~(12) after $\mathcal{M}$ is replaced by $-\mathcal{M}/4$.
Hence the constant $10/3$ in $F \left(x^{}_i\right)$ cannot be eliminated because of the
non-unitarity of $U$, and it will be the dominant part of $F\left(x^{}_i\right)$ contributing
to the total decay amplitude of $\mu^- \to e^- + \gamma$. This point deserves to be
highlighted, so as to distinguish the more phenomenological
{\it minimal unitarity violation} case from a generic
$(3+n)$ active-sterile neutrino mixing case \cite{Fernandez-Martinez:2016lgt}.
	
\item      In view of Refs.~\cite{Inami:1980fz,Lim:1981kv}, where the radiative decays
of quarks are calculated at the one-loop level, we find that a similar loop function can
be extracted after we switch off the contributions of those extra Feynman diagrams
which are irrelevant to Figure~\ref{meg}:
\begin{eqnarray}
\widetilde{F}\left(x^{}_i\right) = - \frac{2x^3_i + 5x^2_i - x^{}_i}{4\left(1-x^{}_i\right)^3}
- \frac{3x^3_i}{2\left(1-x^{}_i\right)^4} \ln x^{}_i \;.
\end{eqnarray}
This result is exactly the loop function of $\mu^- \to e^- + \gamma$ that has been
obtained in Refs.~\cite{Ilakovac:1994kj,Alonso:2012ji,Lindner:2016bgg}.
It is obvious that the results given in Eqs.~(9) and (13) differ from each other
by a constant $-5/6$; namely, $G^{}_\gamma\left(x^{}_i\right) = \widetilde{F}\left(x^{}_i\right)
-5/6$. As discussed above, this constant can be eliminated when the
unitarity of $\mathcal{U}$ is taken into consideration.
\end{itemize}
In short, one should be careful in adopting an explicit expression for the loop function of
radiative decays of charged leptons when discussing the non-unitarity of the PMNS
matrix $U$, because the issue depends on which unitarity violation scheme is under
discussion.
\begin{figure}[t!]
\centering
\includegraphics[width=10cm]{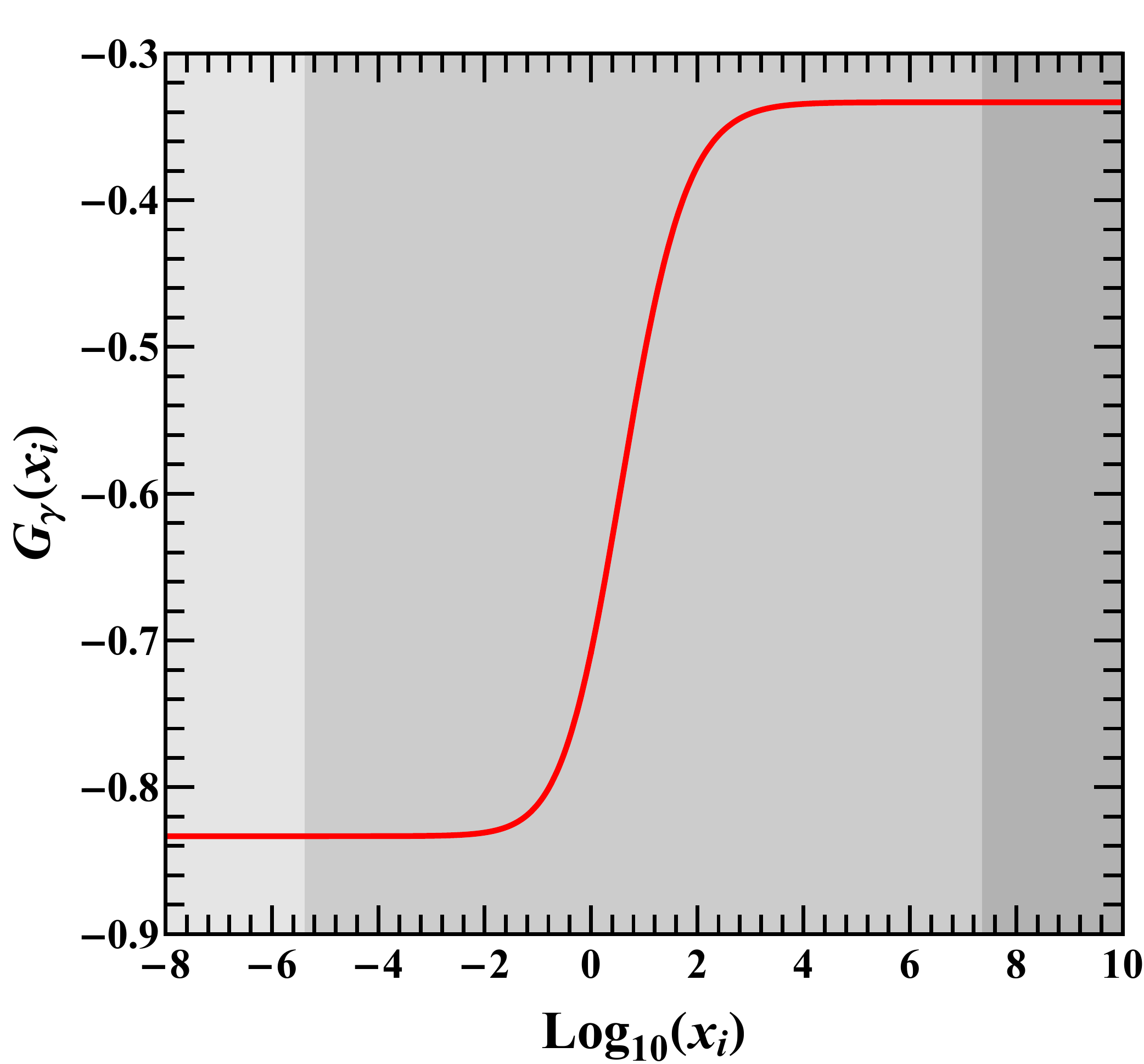}
\caption{A numerical illustration of the loop function $G^{}_\gamma\left(x^{}_i\right)$
evolving with $x^{}_i \equiv \lambda^2_i/M^{2}_W$. In the light gray
($x^{}_i \lesssim 4.0 \times 10^{-6}$) and dark gray ($x^{}_i \gtrsim 2.2 \times 10^7$)
regimes, the asymptotic limits $G^{}_\gamma\left(x^{}_i\right) = -5/6$ and $-1/3$ hold
respectively up to the accuracy of ${\cal O}(10^{-6})$.}
\label{Gr}
\end{figure}

Since the form of $G^{}_\gamma \left(x^{}_i\right)$ itself is universal for both active
and sterile neutrinos, it is very useful to consider its behaviors evolving with $x^{}_i$,
especially in the extreme cases $x^{}_i \ll 1$ (i.e., $\lambda^2_i \ll M^2_W$) and $x^{}_i\gg1$
(i.e., $\lambda^2_i \gg M^2_W$). One can easily observe $G^{}_\gamma \left(x^{}_i\right) =
-5/6$ for $x^{}_i \ll 1$ and $G^{}_\gamma \left(x^{}_i\right) = -1/3$ for $x^{}_i\gg1$,
as numerically illustrated in Figure~\ref{Gr}. In the regions of $x^{}_i \lesssim
4.0 \times 10^{-6}$ (light gray) and $x^{}_i \gtrsim 2.2 \times 10^7$ (dark gray),
or equivalently $\lambda^{}_i \lesssim 0.16$ GeV and
$\lambda^{}_i \gtrsim 3.8 \times 10^5$ GeV, we find that
$G^{}_\gamma \left(x^{}_i\right) + 5/6 < 10^{-6}$ and
$-G^{}_\gamma \left(x^{}_i\right) -1/3 < 10^{-6}$ hold, respectively.
These two thresholds may be used to roughly define the regimes of
``light" and ``heavy" sterile neutrinos when dealing with the contributions of
such new degrees of freedom to the loop function of radiative $\beta^- \to \alpha^- + \gamma$
decays for $(\alpha, \beta) = (e, \mu)$, $(e, \tau)$ and $(\mu, \tau)$.

\begin{figure}[t!]
\centering
\includegraphics[width=10.2cm]{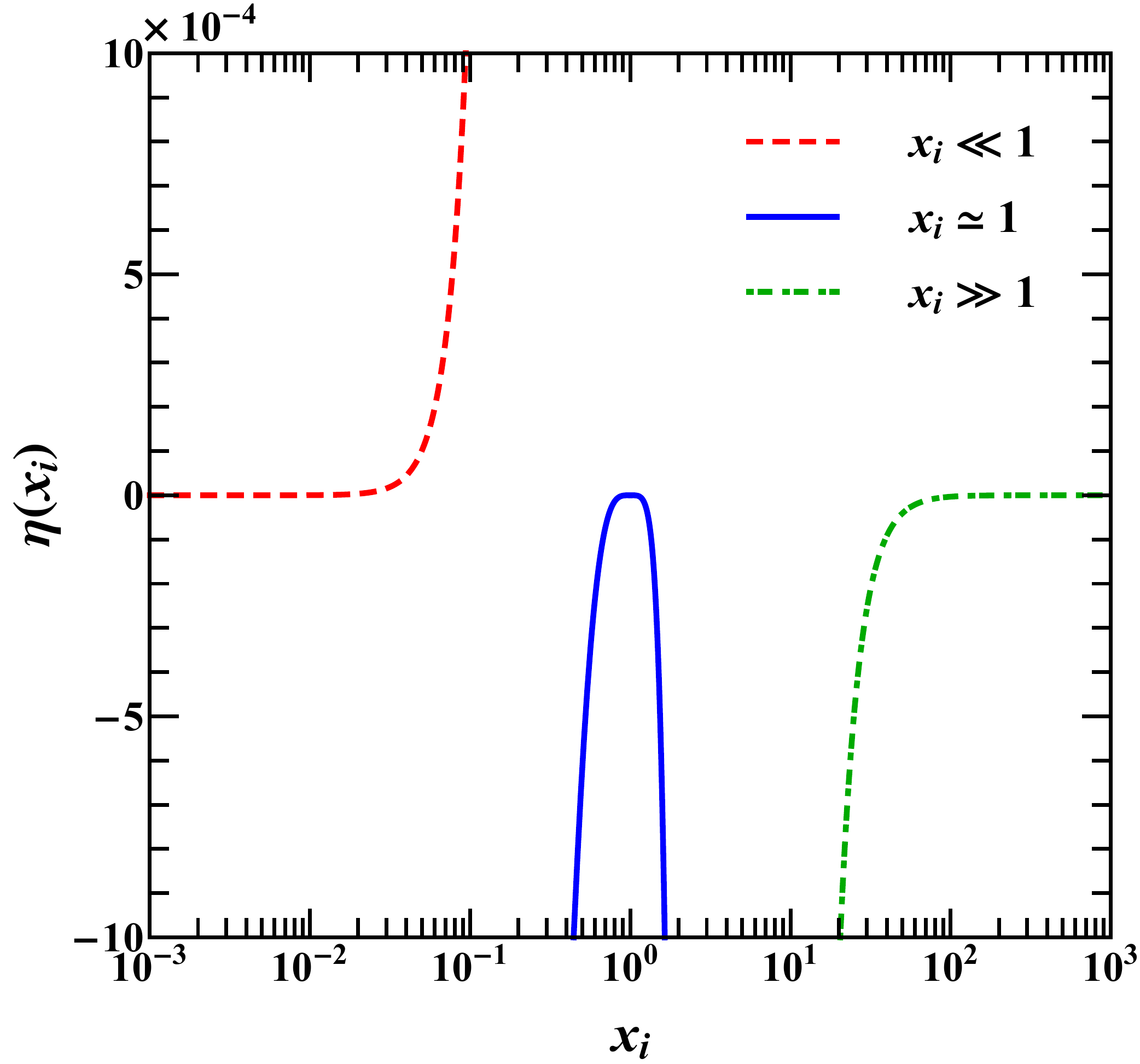}
\caption{A numerical illustration of the relative error $\eta \left(x^{}_i\right)$ between
the exact result of $G^{}_\gamma \left(x^{}_i\right)$ and the analytical approximations
given in Eq.~(14), in which $x^{}_i \lesssim 0.1$, $0.1 \lesssim x^{}_i \lesssim 10$ and
$x^{}_i \gtrsim 10$ have been taken to plot $\eta \left(x^{}_i\right)$ corresponding to the
regimes of $x^{}_i \ll 1$, $x^{}_i \simeq 1$ and $x^{}_i \gg 1$, respectively.}
\label{Grr}
\end{figure}

On the other hand, it is also instructive and helpful to make an analytical approximation
for the expression of $G^{}_\gamma\left(x^{}_i\right)$ in three typical regions of $x^{}_i$
(i.e., $x^{}_i \ll 1$, $x^{}_i \simeq 1$ and $x^{}_i \gg 1$). To a good degree of accuracy,
we obtain the approximate analytical results
\begin{eqnarray}
\widetilde{G}^{}_\gamma \left(x^{}_i\right) = \left\{
\begin{array}{l} \displaystyle -\frac{5}{6} + \frac{x^{}_i}{4} - \frac{x^2_i}{2} -
\frac{x^3_i \left( 11 + 6 \ln x^{}_i \right)}{4} + \mathcal{O}\left(x^4_i\right)
\hspace{1cm} ({\rm for} ~x^{}_i \ll 1) \; ,
\\ \vspace{-0.3cm} \\
\displaystyle - \frac{17}{24} + \frac{3\left( x^{}_i - 1 \right)}{40} -
\frac{\left(x^{}_i - 1 \right)^2}{40} + \frac{3\left(x^{}_i - 1 \right)^3}{280} +
\mathcal{O} \left[ \left( x^{}_i - 1 \right)^4 \right] \hspace{1cm} ({\rm for}
~ x^{}_i \simeq 1) \; ,
\\ \vspace{-0.3cm} \\
\displaystyle -\frac{1}{3} + \frac{11 - 6\ln x^{}_i }{4x^{}_i} + \frac{13 -
	12\ln x^{}_i}{2x^2_i} + \frac{47 - 60 \ln x^{}_i}{4x^3_i} + \mathcal{O}
\left( x^{-4}_i \right)  \hspace{1cm} ({\rm for} ~ x^{}_i\gg1) \; .
\end{array} \right.
\end{eqnarray}
The relative error between such analytical approximations and the exact result of
$G^{}_\gamma (x^{}_i)$, denoted as $\eta (x^{}_i) \equiv \left[\widetilde{G}^{}_\gamma (x^{}_i)
- G^{}_\gamma (x^{}_i)\right]/G^{}_\gamma (x^{}_i)$, is numerically illustrated in Figure~\ref{Grr}, where $x^{}_i \lesssim 0.1$, $0.1 \lesssim x^{}_i \lesssim 10$ and
$x^{}_i \gtrsim 10$ have been taken to plot $\eta \left(x^{}_i\right)$ corresponding to
the ranges of $x^{}_i \ll 1$, $x^{}_i \simeq 1$ and $x^{}_i \gg 1$, respectively.
As can be seen in Figure~\ref{Grr}, $\eta \left(x^{}_i \right) \lesssim 1 \times 10^{-4}$
holds in the regions of $x^{}_i \lesssim 0.05$, $0.67 \lesssim x^{}_i \lesssim 1.36$
and $x^{}_i \gtrsim 38$. This means that in such regions the analytical approximations
made in Eq.~(14) coincide with the exact result of $G^{}_\gamma (x^{}_i)$ very well.
In particular, the masses of heavy Majorana neutrinos in the canonical seesaw mechanism
{\it does} satisfy $x^{}_i \gg 1$.

We proceed to calculate the {\it unpolarized} decay rate of $\mu^- \to e^- + \gamma$ by
using the formula
\begin{eqnarray}
\Gamma \left( \mu^- \to e^- + \gamma \right) = \frac{1}{2m^{}_\mu} \cdot
\frac{1}{8\pi} \left( 1 - \frac{m^2_e}{m^2_\mu} \right) \cdot \frac{1}{2}
\sum \left| \mathcal{M} \right|^2 \; ,
\end{eqnarray}
in which the last part means that $\left| \mathcal{M} \right|^2$ is averaged over the
spin states of $\mu^-$ and summed over the spin states of $e^-$ and the
polarization states of $\gamma$. To be explicit,
\begin{eqnarray}
\frac{1}{2} \sum \left| \mathcal{M} \right| ^2
\hspace{-0.2cm}&=&\hspace{-0.2cm} \displaystyle
- \frac{e^2g^4}{8 \left(4\pi\right)^4 M^4_W} \left| \sum\limits^{3+n}_{i=1} \mathcal{U}^{}_{ei}
\mathcal{U}^\ast_{\mu i} G^{}_\gamma \left(x^{}_i\right) \right|^2 q^{}_\lambda
q^{}_\mu {\rm Tr} \left[ \left( \slashed{p} - \slashed{q} + m^{}_e \right) \sigma^{\rho\lambda}
\left( m^{}_e P^{}_{\rm L} + m^{}_\mu P^{}_{\rm R} \right) \right.
\nonumber \\
\hspace{-0.2cm}&&\hspace{-0.2cm} \times \left. \left( \slashed{p} + m^{}_\mu \right)
\left( m^{}_e P^{}_{\rm R} + m^{}_\mu P^{}_{\rm L} \right)
\sigma^{~\mu}_\rho \right]
\nonumber \\
\hspace{-0.2cm}&=&\hspace{-0.2cm} \frac{e^2g^4}{4 \left(4\pi\right)^4 M^4_W} \left( m^2_\mu +
m^2_e \right) \left( m^2_\mu - m^2_e \right)^2 \left| \sum\limits^{3+n}_{i=1} \mathcal{U}^{}_{ei}
\mathcal{U}^\ast_{\mu i} G^{}_\gamma \left(x^{}_i\right) \right|^2 \; .
\end{eqnarray}
As a result,
\begin{eqnarray}
\Gamma \left( \mu^- \to e^- + \gamma \right) = \frac{\alpha^{}_{\rm em} G^2_{\rm F}
m^5_\mu}{128\pi^4}
\left( 1 + \frac{m^2_e}{m^2_\mu} \right) \left( 1- \frac{m^2_e}{m^2_\mu} \right)^3
\left| \sum\limits^{3+n}_{i=1} \mathcal{U}^{}_{ei} \mathcal{U}^\ast_{\mu i} G^{}_\gamma \left(x^{}_i\right) \right|^2 \;,
\end{eqnarray}
where $\alpha^{}_{\rm em} \equiv e^2/\left(4\pi\right)$ is the fine-structure constant
of electromagnetic interactions, and $G^{}_{\rm F} \equiv g^2/\left(4\sqrt{2} M^2_W\right)$ is
the Fermi coupling constant of weak interactions.

Extending the above calculations to the radiative decays $\tau^- \to e^- + \gamma$ and
$\tau^- \to \mu^- + \gamma$, one may directly write out the unpolarized decay rate of
$\beta^- \to \alpha^- + \gamma$ as follows:
\begin{eqnarray}
\Gamma \left( \beta^- \to \alpha^- + \gamma \right) = \frac{\alpha^{}_{\rm em} G^2_{\rm F}
m^5_\beta}{128\pi^4}
\left( 1 + \frac{m^2_\alpha}{m^2_\beta} \right) \left( 1- \frac{m^2_\alpha}{m^2_\beta} \right)^3
\left| \sum\limits^{3+n}_{i=1} \mathcal{U}^{}_{\alpha i} \mathcal{U}^\ast_{\beta i} G^{}_\gamma \left(x^{}_i\right) \right|^2 \;,
\end{eqnarray}
where $(\alpha, \beta) = (e, \mu)$, $(e, \tau)$ or $(\mu, \tau)$. In comparison, the pure
leptonic decays $\beta^- \to \alpha^- + \overline{\nu}^{}_\alpha + \nu^{}_\beta$ are also mediated
by $W^-$ and their rates in the $(3+n)$ active-sterile neutrino mixing scheme are given by
\begin{eqnarray}
\Gamma \left( \beta^- \to \alpha^- + \overline{\nu}^{}_\alpha + \nu^{}_\beta \right) =
\frac{G^2_{\rm F} m^5_\beta}{192\pi^3}  \left( 1 - 8\frac{m^2_\alpha}{m^2_\beta} \right)
\left[ 1 + \frac{\alpha^{}_{\rm em}}{2\pi} \left( \frac{25}{4} - \pi^2 \right) \right] \sum^{3}_{i=1}
\sum^{3}_{j=1} \left|\mathcal{U}^{}_{\alpha i}\right|^2\left|\mathcal{U}^{}_{\beta j}\right|^2 \; ,
\end{eqnarray}
in which the term proportional to $\alpha^{}_{\rm em}$ stands for the electromagnetic
corrections \cite{Lindner:2016bgg}. Note that
\begin{eqnarray}
\sum^{3}_{i=1} \sum^{3}_{j=1} \left|\mathcal{U}^{}_{\alpha i}\right|^2
\left|\mathcal{U}^{}_{\beta j}\right|^2 =
\sum^{3}_{i=1} \sum^{3}_{j=1} \left|U^{}_{\alpha i}\right|^2
\left|U^{}_{\beta j}\right|^2 = \left(1 - \sum^n_{i=1} \left|R^{}_{\alpha i}\right|^2\right)
\left(1 - \sum^n_{j=1} \left|R^{}_{\beta j}\right|^2\right) \; ,
\end{eqnarray}
where $UU^\dagger + RR^\dagger = I$ has been used, and thus it is expected to be
only slightly departure from one. It is therefore straightforward to
obtain the dimensionless ratio
\begin{eqnarray}
\xi \left( \beta^- \to \alpha^- + \gamma \right)
\hspace{-0.2cm} & \equiv & \hspace{-0.2cm}
\frac{\Gamma \left( \beta^- \to \alpha^- +
\gamma \right)}{\Gamma \left( \beta^- \to \alpha^- + \overline{\nu}^{}_\alpha + \nu^{}_\beta \right)}
\nonumber \\
\hspace{-0.2cm} & \simeq & \hspace{-0.2cm}
\frac{3\alpha^{}_{\rm em}}{2\pi} \left|
\sum^{3}_{i=1} U^{}_{\alpha i} U^\ast_{\beta i} G^{}_\gamma \left( \frac{m^2_i}{M^2_W}
\right) + \sum^{n}_{i=1} R^{}_{\alpha i} R^\ast_{\beta i} G^{}_\gamma \left( \frac{M^2_i}{M^2_W}
\right) \right|^2 \; , \hspace{0.6cm}
\end{eqnarray}
in the leading-order approximation by neglecting those next-to-leading-order and higher-order
contributions. Taking account of the asymptotic behaviors of the loop function shown above,
we are going to discuss how to constrain the unitarity polygons by means of the experimental
upper bounds of $\xi \left( \beta^- \to \alpha^- + \gamma \right)$.

\section{Constraints on the unitarity polygons}

In the $(3+n)$ active-sterile neutrino mixing scheme, it is natural to assume that the $n$ sterile
neutrinos are heavy enough such that the seesaw mechanism takes effect to explain why three
active neutrinos are so light. Given this well-motivated picture with $M^{}_i \gg M^{}_W$,
one may accordingly simplify the loop functions $G^{}_\gamma\left(m^2_i/M^2_W\right)$ and
$G^{}_\gamma\left(M^2_i/M^2_W\right)$. Taking account of the analytical approximations made
in Eq.~(14) and the unitarity condition $UU^\dagger + RR^\dagger = I$, we obtain
\begin{eqnarray}
\xi \left( \beta^- \to \alpha^- + \gamma \right)
\hspace{-0.2cm} & \simeq & \hspace{-0.2cm}
\frac{3\alpha^{}_{\rm em}}{2\pi} \left|
\sum^{3}_{i=1} U^{}_{\alpha i} U^\ast_{\beta i} \left(-\frac{5}{6} + \frac{1}{4}\cdot
\frac{m^2_i}{M^2_W}\right) -\frac{1}{3} \sum^{n}_{i=1} R^{}_{\alpha i} R^\ast_{\beta i} \right|^2
\nonumber \\
\hspace{-0.2cm} & = & \hspace{-0.2cm}
\frac{3\alpha^{}_{\rm em}}{8\pi} \left|
\sum^{3}_{i=1} U^{}_{\alpha i} U^\ast_{\beta i} \left(1 - \frac{1}{2}\cdot
\frac{m^2_i}{M^2_W}\right) \right|^2 \; ,
\end{eqnarray}
up to the leading order of $R^{}_{\alpha i} R^*_{\beta i}$ (for $i=1, 2, \cdots, n$).
Two immediate observations are in order.
\begin{itemize}
\item     If all the new degrees of freedom are switched off, the PMNS matrix $U$ is
exactly unitary and thus one arrives at the ``standard" result
\begin{eqnarray}
\xi \left( \beta^- \to \alpha^- + \gamma \right) \simeq
\frac{3\alpha^{}_{\rm em}}{32\pi} \left|
\sum^{3}_{i=1} U^{}_{\alpha i} U^\ast_{\beta i} \frac{m^2_i}{M^2_W}\right|^2
= \frac{3\alpha^{}_{\rm em}}{32\pi} \left|
\sum^{3}_{i=2} U^{}_{\alpha i} U^\ast_{\beta i} \frac{\Delta m^2_{i1}}{M^2_W}\right|^2
\lesssim {\cal O}(10^{-54}) \; ,
\end{eqnarray}
where $\Delta m^2_{i1} \equiv m^2_i - m^2_1$ is defined (for $i = 2, 3$), and current neutrino
oscillation data \cite{Zyla:2020zbs} have been taken into account in making the above estimate
\cite{Xing:2019vks}.
In this case it is therefore hopeless to measure any radiative decays of charged leptons.

\item     If the PMNS matrix $U$ is not exactly unitary due to the existence of
slight mixing between three
active neutrinos and $n$ sterile heavy neutrinos as discussed above, then it is quite safe to
neglect the $m^2_i/M^{2}_W$ terms in Eq.~(22) and obtain a straightforward but strong constraint on
the {\it effective} apex of a given {\it effective} Dirac-type unitarity triangle shown
in Figure~\ref{triangle}:
\begin{eqnarray}
\left| \sum^{3}_{i=1} U^{}_{\alpha i} U^\ast_{\beta i} \right| =
\left| \sum^{n}_{i=1} R^{}_{\alpha i} R^\ast_{\beta i} \right| =
\sqrt{\frac{8\pi}{3 \alpha^{}_{\rm em}} \xi \left( \beta^- \to \alpha^- + \gamma \right)} \
\simeq 33.88 \sqrt{\xi \left( \beta^- \to \alpha^- + \gamma \right)} \; ,
\end{eqnarray}
where $\alpha^{}_{\rm em} \simeq 1/137$ has been input at low energies.
\end{itemize}
So Eq.~(24) provides us with a realistic way to probe or constrain the seesaw-induced
unitarity violation in radiative $\beta^- \to \alpha^- + \gamma$ decays (see also
Ref.~\cite{Fernandez-Martinez:2016lgt} for some similar discussions and observations).
It is worth remarking that the coefficient in front of $\sqrt{\xi \left( \beta^- \to
\alpha^- + \gamma \right)}$ under discussion differs from that obtained in the
{\it minimal unitarity violation} scheme by a factor $5/3$ (i.e., the latter amounts
to $\sqrt{24\pi/\left(25 \alpha^{}_{\rm em}\right)} \simeq 20.33$
\cite{Antusch:2006vwa,Antusch:2014woa,Calibbi:2017uvl}).

To illustrate, let us take into account current experimental upper bounds on the
branching fractions of $\beta^- \to \alpha^- + \gamma$ decays
and $\beta^- \to \alpha^- + \overline{\nu}^{}_\alpha + \nu^{}_\beta$ decays \cite{Zyla:2020zbs}.
That is,
\begin{eqnarray}
{\cal B}\left(\mu^- \to e^- + \gamma\right) \hspace{-0.2cm} & < & \hspace{-0.2cm}
4.2 \times 10^{-13} \; ,
\nonumber \\
{\cal B}\left(\tau^- \to e^- + \gamma\right) \hspace{-0.2cm} & < & \hspace{-0.2cm}
3.3 \times 10^{-8} \; ,
\nonumber \\
{\cal B}\left(\tau^- \to \mu^- + \gamma\right) \hspace{-0.2cm} & < & \hspace{-0.2cm}
4.4 \times 10^{-8} \; ,
\end{eqnarray}
at the $90\%$ confidence level, together with
${\cal B}\left(\mu^- \to e^- + \overline{\nu}^{}_e + \nu^{}_\mu\right)
\simeq 100\%$, ${\cal B}\left(\tau^- \to e^- + \overline{\nu}^{}_e + \nu^{}_\tau\right)
\simeq 17.82\%$ and ${\cal B}\left(\tau^- \to \mu^- + \overline{\nu}^{}_\mu + \nu^{}_\tau\right)
\simeq 17.39\%$. Then we obtain the ratios
\begin{eqnarray}
\xi\left(\mu^- \to e^- + \gamma\right) \hspace{-0.2cm} & < & \hspace{-0.2cm}
4.20 \times 10^{-13} \; ,
\nonumber \\
\xi\left(\tau^- \to e^- + \gamma\right) \hspace{-0.2cm} & < & \hspace{-0.2cm}
1.85 \times 10^{-7} \; ,
\nonumber \\
\xi\left(\tau^- \to \mu^- + \gamma\right) \hspace{-0.2cm} & < & \hspace{-0.2cm}
2.53 \times 10^{-7} \; .
\end{eqnarray}
A combination of Eqs.~(24) and (26) leads us to the constraints
\begin{eqnarray}
\left| \sum^{3}_{i=1} U^{}_{e i} U^\ast_{\mu i} \right| =
\left| \sum^{n}_{i=1} R^{}_{e i} R^\ast_{\mu i} \right|
\hspace{-0.2cm} & < & \hspace{-0.2cm}
2.20 \times 10^{-5} \; ,
\nonumber \\
\left| \sum^{3}_{i=1} U^{}_{e i} U^\ast_{\tau i} \right| =
\left| \sum^{n}_{i=1} R^{}_{e i} R^\ast_{\tau i} \right|
\hspace{-0.2cm} & < & \hspace{-0.2cm}
1.46 \times 10^{-2} \; ,
\nonumber \\
\left| \sum^{3}_{i=1} U^{}_{\mu i} U^\ast_{\tau i} \right| =
\left| \sum^{n}_{i=1} R^{}_{\mu i} R^\ast_{\tau i} \right|
\hspace{-0.2cm} & < & \hspace{-0.2cm}
1.70 \times 10^{-2} \; .
\end{eqnarray}
These results clearly show that the unitarity polygons in Figure~\ref{triangle}
can be treated as the effective unitarity triangles, since their differences are at most at
the ${\cal O}(10^{-2})$ level.

Now we take a brief look at the uncertainties induced by the approximations made to obtain Eq.~(24). If the next-to-leading-order terms of $G^{}_\gamma\left(m^2_i/M^2_W\right)$ and
$G^{}_\gamma\left(M^2_i/M^2_W\right)$ are both taken into account, Eq.~(22) will be
replaced with
\begin{eqnarray}
\xi \left( \beta^- \to \alpha^- + \gamma \right)
\hspace{-0.2cm} & \simeq & \hspace{-0.2cm}
\frac{3\alpha^{}_{\rm em}}{2\pi}
\left| \sum^{3}_{i=1} U^{}_{\alpha i} U^\ast_{\beta i} \left(-\frac{5}{6} + \frac{1}{4}\cdot
\frac{m^2_i}{M^2_W}\right) \right.
\nonumber \\
\hspace{-0.2cm} & & \hspace{-0.2cm}
+ \left. \sum^{n}_{i=1} R^{}_{\alpha i} R^\ast_{\beta i} \left[ - \frac{1}{3} +
\frac{1}{4}\cdot \frac{M^2_W}{M^2_i} \left(11 + 6\ln \frac{M^2_W}{M^2_i}\right)\right]\right|^2
\nonumber \\
\hspace{-0.2cm} & \simeq & \hspace{-0.2cm}
\frac{3\alpha^{}_{\rm em}}{8\pi}
\left|\sum^{3}_{i=1} U^{}_{\alpha i} U^\ast_{\beta i} - \frac{1}{2} \left[\sum^{3}_{i=1}
U^{}_{\alpha i} U^\ast_{\beta i} \frac{m^2_i}{M^2_W} \right.\right.
\nonumber \\
\hspace{-0.2cm} & & \hspace{-0.2cm}
+ \left.\left. \sum^{n}_{i=1} R^{}_{\alpha i} R^\ast_{\beta i}
\frac{M^2_W}{M^2_i} \left(11 + 6\ln \frac{M^2_W}{M^2_i}\right)\right]\right|^2 \;.
\end{eqnarray}
Making use of Eq.~(28) and the formula $\left|\left|a\right|-\left|b\right|\right| 
\leq \left|a+b\right| \leq \left|a\right| + \left|b\right|$ 
for two arbitrary complex quantities $a$ and $b$, one may
estimate small corrections to the leading-order result given in Eq.~(24):
\begin{eqnarray}
\left| \sum^{3}_{i=1} U^{}_{\alpha i} U^\ast_{\beta i} \right|
\hspace{-0.2cm} & \simeq & \hspace{-0.2cm}
33.88 \sqrt{\xi \left( \beta^- \to \alpha^- + \gamma \right)} \pm \frac{1}{2}
\left[ \sum^{3}_{i=1} \left|  U^{}_{\alpha i} U^\ast_{\beta i}
\frac{m^2_i}{M^2_W} \right| \right.
\nonumber \\
\hspace{-0.2cm} & & \hspace{-0.2cm}
+ \left. \sum^{n}_{i=1} \left| R^{}_{\alpha i} R^\ast_{\beta i}
\frac{M^2_W}{M^2_i} \left(11 + 6\ln \frac{M^2_W}{M^2_i}\right)\right|\right] \; ,
\end{eqnarray}
where the next-to-leading-order contributions have been treated as the uncertainties.
Taking account of $U^{}_{\alpha i} \sim 1$, $R^{}_{\alpha i} \sim 10^{-2}$ and
$m^{}_i \sim 1$ eV for example, we find that the terms in the squre bracket of Eq.~(29)
are roughly at the level of $7.2 \times 10^{-9}$ for $M^{}_i \sim 100$ TeV or
$1.9 \times 10^{-5}$ for $M^{}_i \sim 1$ TeV. So the
leading-order approximation made in Eq.~(24) and its intriguing consequence obtained 
in Eq.~(27) are actually safe enough for a natural
seesaw mechanism with $M^{}_i \gtrsim 100$ TeV. The latter is essentially 
the defined regime of ``heavy" sterile neutrinos in section 2. 
If $M^{}_i \sim 1$ TeV holds, however, the upper bound on 
$\left| \sum\limits^{3}_{i=1} U^{}_{e i} U^\ast_{\mu i} \right|$ achieved
in Eq.~(27) will suffer a large uncertainty induced by the next-to-leading-order
contribution of $G^{}_\gamma (M^2_i/M^2_W)$, but the upper limits on 
$\left| \sum\limits^{3}_{i=1} U^{}_{e i} U^\ast_{\tau i} \right|$ and
$\left| \sum\limits^{3}_{i=1} U^{}_{\mu i} U^\ast_{\tau i} \right|$ remain valid. 
\begin{figure}[t!]
	\centering
	\includegraphics[width=1\linewidth]{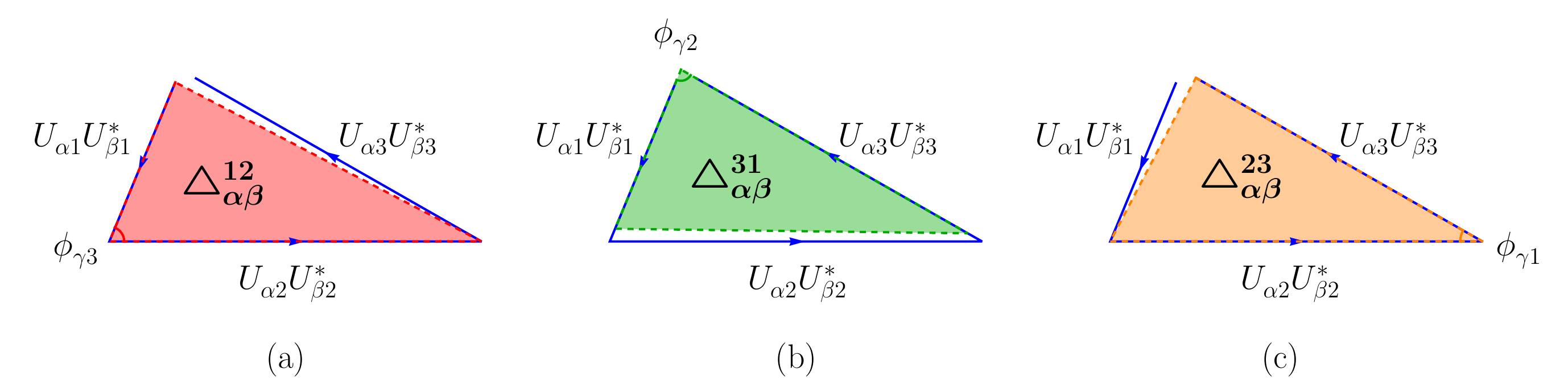}
	\caption{Three sub-triangles $\triangle^{ij}_{\alpha\beta}$ of a given unitarity polygon
with definite $\alpha$ and $\beta$ flavor indices, formed by two
vectors $U^{}_{\alpha i} U^*_{\beta i}$ and $U^{}_{\alpha j} U^*_{\beta j}$ (for $i \neq j$)
in the complex plane.}
	\label{tri}
\end{figure}

The future precision measurements at low energies will allow us to test unitarity of
the $3\times 3$ PMNS matrix $U$ to a much better degree of accuracy. In this
connection it makes sense to look at the geometry of a given unitarity polygon as
compared with the geometry of its three sub-triangles $\triangle^{ij}_{\alpha\beta}$ formed by
two vectors $U^{}_{\alpha i} U^*_{\beta i}$ and $U^{}_{\alpha j} U^*_{\beta j}$ (for $i \neq j$)
in the complex plane, as illustrated by Figure~\ref{tri}. The inner intersection
angle between these two vectors can therefore be defined as
\begin{eqnarray}
\phi^{}_{\gamma 1} \equiv \arg \left( -\frac{U^{}_{\alpha 2} U^\ast_{\beta 2}}{U^{}_{\alpha 3} U^\ast_{\beta 3}} \right) \;,
\quad
\phi^{}_{\gamma 2} \equiv \arg \left( -\frac{U^{}_{\alpha 3} U^\ast_{\beta 3}}{U^{}_{\alpha 1} U^\ast_{\beta 1}} \right) \;,
\quad
\phi^{}_{\gamma 3} \equiv \arg \left( -\frac{U^{}_{\alpha 1} U^\ast_{\beta 1}}{U^{}_{\alpha 2} U^\ast_{\beta 2}} \right) \;,
\end{eqnarray}
where $(\alpha, \beta, \gamma) = (e, \mu, \tau)$, $(\mu, \tau, e)$ or $(\tau, e, \mu)$. It is
obvious that the sum
\begin{eqnarray}
\phi^{}_{\gamma 1} + \phi^{}_{\gamma 2} + \phi^{}_{\gamma 3} = \pi \;
\end{eqnarray}
holds by definition, even though $U^{}_{\alpha 1} U^\ast_{\beta 1} + U^{}_{\alpha 2} U^\ast_{\beta 2}
+ U^{}_{\alpha 3} U^\ast_{\beta 3} \neq 0$. This point is sometimes overlooked in the quark sector
when discussing the CKM unitarity test.

One may establish a direct link between the areas of those sub-triangles
$\triangle^{ij}_{\alpha\beta}$, denoted as $S^{ij}_{\alpha\beta}$,
and the Jarlskog-like invariants which are defined in Eq.~(4) and satisfy
the relation $\mathcal{J}^{ij}_{\alpha\beta} = \mathcal{J}^{ji}_{\beta\alpha} =
-\mathcal{J}^{ji}_{\alpha\beta} = - \mathcal{J}^{ij}_{\beta\alpha}$. Namely,
we have
\begin{eqnarray}
S^{ij}_{\alpha \beta} = \frac{1}{2} \left| \mathcal{J}^{ij}_{\alpha\beta} \right| \; ,
\end{eqnarray}
where $(\alpha, \beta) = (e, \mu)$, $(\mu, \tau)$
or $(\tau, e)$ and $(i, j) = (1, 2)$, $(2, 3)$ or $(3, 1)$.
If $R = 0$ and $U$ is exactly unitary, all the possible Jarlskog-like invariants
are identical in magnitude \cite{Jarlskog:1985ht}. This will not be true anymore in
the existence of active-sterile neutrino mixing, implying that the areas of
$\triangle^{ij}_{\alpha\beta}$ are not exactly equal. To illustrate this point in a
more transparent way, we consider the $(3+3)$ active-sterile neutrino mixing scenario
and write out the explicit expressions of $U$ and $R$ in an Euler-like parametrization
of the $6\times 6$ unitary matrix $\cal U$, as outlined in Appendix C.
As a result, we find
\begin{eqnarray}
\left| S^{12}_{e\mu} - S^{23}_{e\mu} \right| \hspace{-0.2cm} & = & \hspace{-0.2cm} \frac{1}{2}
\left| s^{}_{12} c^{}_{13} \left( c^{}_{12} c^{}_{23} {\rm Im} X^{}_1 - s^{}_{12} s^{}_{13} s^{}_{23}
{\rm Im} X^{}_2 \right) \right| \;,
\nonumber \\
\left| S^{12}_{e\mu} - S^{31}_{e\mu} \right| \hspace{-0.2cm} & = & \hspace{-0.2cm} \frac{1}{2}
\left| c^{}_{12} c^{}_{13} \left( s^{}_{12} c^{}_{23} {\rm Im} X^{}_1 + c^{}_{12} s^{}_{13} s^{}_{23}
{\rm Im} X^{}_2  \right)  \right| \;,
\nonumber \\
\left| S^{23}_{e\mu} - S^{31}_{e\mu} \right| \hspace{-0.2cm} & = & \hspace{-0.2cm} \frac{1}{2}
\left| c^{}_{13} s^{}_{13} s^{}_{23} {\rm Im} X^{}_2 \right| \; ;
\end{eqnarray}
and
\begin{eqnarray}
\left| S^{12}_{\tau e} - S^{23}_{\tau e} \right| \hspace{-0.2cm} & = & \hspace{-0.2cm} \frac{1}{2}
\left| s^{}_{12} c^{}_{13} \left( c^{}_{12} s^{}_{23} {\rm Im} Y^{}_1 + s^{}_{12} s^{}_{13} c^{}_{23}
{\rm Im} Y^{}_2 \right) \right| \;,
\nonumber \\
\left| S^{12}_{\tau e} - S^{31}_{\tau e} \right| \hspace{-0.2cm} & = & \hspace{-0.2cm} \frac{1}{2}
\left| c^{}_{12} c^{}_{13} \left( s^{}_{12} s^{}_{23} {\rm Im} Y^{}_1 - c^{}_{12} s^{}_{13} c^{}_{23}
{\rm Im} Y^{}_2  \right)  \right| \;,
\nonumber \\
\left| S^{23}_{\tau e} - S^{31}_{\tau e} \right| \hspace{-0.2cm} & = & \hspace{-0.2cm} \frac{1}{2}
\left| c^{}_{13} s^{}_{13} c^{}_{23} {\rm Im} Y^{}_2 \right| \; ;
\end{eqnarray}
as well as
\begin{eqnarray}
\left| S^{12}_{\mu\tau} - S^{23}_{\mu\tau} \right| \hspace{-0.2cm} & = & \hspace{-0.2cm} \frac{1}{2}
\left| \left( c^2_{12} - s^2_{12} s^2_{13} \right) c^{}_{23} s^{}_{23} {\rm Im} Z^{}_1 +
c^{}_{12} s^{}_{12} s^{}_{13} \left(c^2_{23} {\rm Im} Z^{}_2 - s^2_{23} {\rm Im} Z^{}_3 \right) \right| \;,
\nonumber \\
\left| S^{12}_{\mu\tau} - S^{31}_{\mu\tau} \right| \hspace{-0.2cm} & = & \hspace{-0.2cm} \frac{1}{2}
\left| \left( c^2_{12} s^2_{13} - s^2_{12} \right) c^{}_{23} s^{}_{23} {\rm Im} Z^{}_1 + c^{}_{12}
s^{}_{12} s^{}_{13} \left(c^2_{23} {\rm Im} Z^{}_2 - s^2_{23} {\rm Im} Z^{}_3 \right) \right| \;,
\nonumber \\
\left| S^{23}_{\mu\tau} - S^{31}_{\mu\tau} \right| \hspace{-0.2cm} & = & \hspace{-0.2cm} \frac{1}{2}
\left| c^2_{13} c^{}_{23} s^{}_{23} {\rm Im} Z^{}_1 \right| \;,
\end{eqnarray}
where $c^{}_{ij} \equiv \cos \theta^{}_{ij}$, $s^{}_{ij} \equiv \sin \theta^{}_{ij}$ (for $ij=12, 13, 23$),
and $X^{}_i$, $Y^{}_i$ and $Z^{}_i$ (for $i=1, 2, 3$) are defined in Eq.~(C9) of Appendix C
and depend on the active-sterile neutrino mixing angles and CP-violating phases.
Eqs.~(C6)---(C8) tell us that all the nine Jarlskog-like invariants $\mathcal{J}^{ij}_{\alpha\beta}$
will be reduced to the unique Jarlskog invariant  $\mathcal{J}^{}_0 = \sin 2\theta^{}_{12}
\sin 2\theta^{}_{13} \cos\theta^{}_{13} \sin 2\theta^{}_{23} \sin\delta/8$
in the standard parametrization of $U$ if $U$ is exactly unitary,
and in this case triangles $\triangle^{12}_{\alpha\beta}$, $\triangle^{23}_{\alpha\beta}$ and
$\triangle^{31}_{\alpha\beta}$ are equivalent to one another and their areas are all equal to
$\left|{\cal J}^{}_0 \right|/2$. Since $X^{}_i$, $Y^{}_i$ and $Z^{}_i$ (for $i=1, 2, 3$) are all of
${\cal O}\left(\sin^{2}\theta^{}_{ij}\right)$ with $\theta^{}_{ij} \lesssim {\cal O}\left(0.1\right)$
being the small active-sterile neutrino mixing angles (for $i=1,2,3$ and $j=4,5,6$), the areas
of three sub-triangles of a given unitarity hexagon are expected to be different from one another
at most at the level of one percent or much smaller. Such an observation is certainly true
for a generic unitarity polygon, and that is why the latter can be reduced to an effective
unitarity triangle with an effective apex as illustrated by Figure~\ref{triangle}.

Here it is worth mentioning that the areas of triangles $\triangle^{12}_{\alpha\beta}$,
$\triangle^{23}_{\alpha\beta}$ and $\triangle^{31}_{\alpha\beta}$ can in principle be
measured in some long-baseline neutrino oscillation experiments. To see this point,
let us assume all the sterile particles in our $(3+n)$ active-sterile neutrino mixing
scheme to be kinematically forbidden and write out
the possibilities of active $\nu^{}_\alpha \to \nu^{}_\beta$ oscillations in
vacuum \cite{Xing:2007zj}:
\begin{eqnarray}
P \left( \nu^{}_\alpha \to \nu^{}_\beta \right) = \displaystyle \frac{\sum\limits^{3}_{i=1}
\left| U^{}_{\alpha i} \right|^2 \left| U^{}_{\beta i} \right|^2 + 2 \sum\limits^{}_{i<j}
{\rm Re} \left( U^{}_{\alpha i} U^{}_{\beta j} U^\ast_{\alpha j} U^\ast_{\beta i} \right)
\cos \Delta^{}_{ij} - 2 \sum\limits^{}_{i<j} \mathcal{J}^{ij}_{\alpha\beta}
\sin \Delta^{}_{ij} }{ \left( UU^\dagger \right)^{}_{\alpha\alpha} \left( UU^\dagger
\right)^{}_{\beta\beta} } \;,
\end{eqnarray}
where $\Delta^{}_{ij} \equiv \left(m^2_i - m^2_j\right) L/\left(2E\right)$.
The possibilities of $\overline{\nu}^{}_\alpha \to \overline{\nu}^{}_\beta$ oscillations
in vacuum can be easily achieved from Eq.~(36) by making the replacement $U \to U^\ast$.
Then the CP-violating asymmetries between $\nu^{}_\alpha \to \nu^{}_\beta$ and
$\overline{\nu}^{}_\alpha \to \overline{\nu}^{}_\beta$ oscillations are directly determined
by the relevant Jarlskog-like invariants as follows:
\begin{eqnarray}
\mathcal{A}^{}_{\alpha\beta} \hspace{-0.2cm} & \equiv & \hspace{-0.2cm} P \left( \nu^{}_\alpha
\to \nu^{}_\beta \right) - P \left( \overline{\nu}^{}_\alpha \to \overline{\nu}^{}_\beta \right)
= - \frac{ 4 \sum\limits^{}_{i<j} \mathcal{J}^{ij}_{\alpha\beta} \sin \Delta^{}_{ij} }
{\left( UU^\dagger \right)^{}_{\alpha\alpha} \left( UU^\dagger \right)^{}_{\beta\beta}}
\nonumber \\
\hspace{-0.2cm} & \simeq & \hspace{-0.2cm} - 4 \sum\limits^{}_{i<j} \left[ 1 +
\left( RR^\dagger \right)^{}_{\alpha\alpha} +  \left( RR^\dagger \right)^{}_{\beta\beta} \right]
\mathcal{J}^{ij}_{\alpha\beta} \sin \Delta^{}_{ij} \; ,
\end{eqnarray}
in which the small active-sterile neutrino mixing effects hidden in $R$ have been taken
into account. Note that the $\left( RR^\dagger \right)^{}_{\alpha\alpha} +
\left( RR^\dagger \right)^{}_{\beta\beta}$ term will cancel the $\mathcal{X}^\prime$,
$\mathcal{Y}^\prime$ or $\mathcal{Z}^\prime$ term in the expressions of $\mathcal{J}^{ij}_{\alpha\beta}$
as shown in Eqs.~(C6)---(C10), but the combinations
$\left[ 1 + \left( RR^\dagger \right)^{}_{\alpha\alpha} +  \left( RR^\dagger \right)^{}_{\beta\beta}
\right] \mathcal{J}^{ij}_{\alpha\beta}$ are still different from
${\cal J}^{}_0$ because of $R \neq 0$. Of course, one has to take into account the terrestrial
matter effects on $\nu^{}_\alpha \to \nu^{}_\beta$ and
$\overline{\nu}^{}_\alpha \to \overline{\nu}^{}_\beta$ oscillations in a realistic
long-baseline experiment \cite{Xing:2007zj,Goswami:2008mi}. Moreover, the detector efficiencies
and the associated systematics should be taken into consideration to probe the tiny $R$-induced
corrections to ${\cal A}^{}_{\alpha\beta}$. All such and other possible
uncertainties are much larger than the strength of $R$-induced CP violation
in current experimental environments, and hence
it will be extremely difficult (if not impossible) to establish a convincing signal of
this kind of new physics even in the foreseeable future.

Finally, let us make some brief comments on the situation that the sterile neutrinos are
light enough such that $M^{}_i \ll M^{}_W$ holds (for $i = 1, 2, \cdots, n$)
\cite{Drewes:2013gca}. In this case, the result of $\xi\left(\beta^- \to \alpha^- +
\gamma\right)$ obtained in Eq.~(21) can be simplified to
\begin{eqnarray}
\xi\left(\beta^- \to \alpha^- + \gamma\right)
\simeq \frac{3\alpha^{}_{\rm em}}{32\pi M^4_W}
\left| \left< M \right>^{2}_{\alpha\beta} \right|^2 \;,
\end{eqnarray}
where the effective neutrino mass $\left< M \right>^{}_{\alpha\beta}$ are defined as
\begin{eqnarray}
\left< M \right>^{2}_{\alpha\beta} = \sum^{3}_{i=1} m^2_i U^{}_{\alpha i} U^\ast_{\beta i}
+ \sum^{n}_{i=1} M^2_i R^{}_{\alpha i} R^\ast_{\beta i} \;.
\end{eqnarray}
It is obvious that $\left< M \right>^{}_{\alpha\beta}$ is insensitive to the
Majorana phases hidden in $U$ and $R$, and thus it is closely related to lepton
flavor violation in the charged-lepton sector. Taking account of current experimental
bounds on $\xi\left(\beta^- \to \alpha^- + \gamma\right)$ given in Eq.~(26),
we immediately arrive at
\begin{eqnarray}
\left| \left< M \right>^{}_{e \mu} \right| < 0.53 ~{\rm GeV} \;,\quad
\left| \left< M \right>^{}_{e \tau} \right| < 13.73 ~{\rm GeV} \;,\quad
\left| \left< M \right>^{}_{\mu \tau}
\right| < 14.84 ~{\rm GeV} \;.
\end{eqnarray}
Such upper limits remain too large to really probe the magnitudes of $m^{}_i$ and $M^{}_i$
in $\left< M \right>^{}_{\alpha\beta}$.

\section{Summary}

We are entering the era of precision measurements of both flavor oscillations of massive
neutrinos and lepton flavor violation in the charged-lepton sector. A burning issue is to test
unitarity of the $3\times 3$ PMNS matrix $U$ so as to probe or constrain possible
new but sterile degrees of freedom which may slightly mix with three active neutrino
species. The most popular example of this kind is the heavy Majorana neutrinos in the
canonical seesaw mechanism, although much lighter sterile neutrinos are also taken into account
in some low-scale seesaw models or purely from a phenomenological point of view. In this
connection the radiative decays of charged leptons in the form of $\beta^- \to \alpha^- + \gamma$,
which may take place via both active and sterile neutrinos in the one-loop,
are expected to be an ideal tool to examine the departure of a unitarity polygon from the standard
unitarity triangle of $U$ in the $(3+n)$ active-sterile neutrino mixing scheme.

That is why we have calculated the rates of radiative $\beta^- \to \alpha^- + \gamma$ decays
in the {\it unitary} gauge, and confirmed the results obtained previously in
Refs.~\cite{Ilakovac:1994kj,Alonso:2012ji}
\footnote{Throughout this work we have focused on the non-supersymmetric active-sterile
neutrino mixing scheme. We refer the reader to Refs.~\cite{Hisano:1995cp,Hisano:1995nq}
for charged lepton flavor violation in the supersymmetric seesaw scenarios.}.
We have made it clear that constraints on the unitarity of $U$ extracted from
$\beta^- \to \alpha^- + \gamma$ decays in the {\it minimal unitarity violation}
scheme differ from those obtained in the canonical seesaw mechanism with $n$ heavy Majorana
neutrinos by a factor $5/3$. In such a natural seesaw case we have demonstrated that the rates
of $\beta^- \to \alpha^- + \gamma$ can be used to cleanly and strongly constrain the effective
apex of a unitarity polygon as shown in Figure~\ref{triangle}, and discussed its geometry
as compared with the geometry of its three sub-triangles $\triangle^{ij}_{\alpha\beta}$ formed
by two vectors $U^{}_{\alpha i} U^*_{\beta i}$ and $U^{}_{\alpha j} U^*_{\beta j}$
(for $i \neq j$) in the complex plane. It is found that the areas of such sub-triangles can be
described in terms of the Jarlskog-like invariants of CP violation
${\cal J}^{ij}_{\alpha\beta}$, and their small differences may serve as a signal of slight
unitarity violation of the $3\times 3$ PMNS matrix $U$. These observations should be useful
to test the unitarity of $U$ when more accurate experimental data are available in the
foreseeable future, and they can certainly be extended to those simplified seesaw cases
(e.g., the minimal seesaw scenarios \cite{Xing:2020ald}) with fewer free parameters.

\section*{Acknowledgements}

We would like to thank Enrique Fernandez-Martinez, Zhi-cheng Liu and Shun Zhou for
very useful discussions. This work was supported in part by the National Natural
Science Foundation of China under grant No. 11775231 and grant No. 11835013.

\newpage

\begin{flushleft}
{\Large\bf Appendices}
\end{flushleft}

\appendix

\renewcommand{\theequation}{\thesection\arabic{equation}}
\setcounter{equation}{0}

\section{Dimensional regularization}

In $D \equiv 4 - \varepsilon$ dimensions, the Dirac matrices satisfy
\begin{eqnarray}
\left\{ \gamma^\mu, \gamma^\nu \right\} = 2 g^{\mu\nu} \;,
\end{eqnarray}
and the Minkowski metric tensor $g^{\mu\nu}$ satisfies
\begin{eqnarray}
g^{\mu\nu} \hspace{-0.2cm}&=&\hspace{-0.2cm} g^{\nu\mu} \;,
\nonumber
\\
g^{\mu\rho}g^{~\nu}_{\rho} \hspace{-0.2cm}&=&\hspace{-0.2cm} g^{\mu\nu} \;,
\nonumber
\\
g^{\mu\nu} g^{}_{\mu\nu} \hspace{-0.2cm}&=&\hspace{-0.2cm} D \;.
\end{eqnarray}
Given Eqs.~(A1) and (A2), one may derive
\begin{eqnarray}
\gamma^\mu \gamma^{}_\mu \hspace{-0.2cm}&=&\hspace{-0.2cm} D \;,
\nonumber
\\
\gamma^\mu \gamma^\nu \gamma^{}_\mu \hspace{-0.2cm}&=&\hspace{-0.2cm} -
\left( D-2 \right) \gamma^\nu \;,
\nonumber
\\
\gamma^\mu \gamma^\nu \gamma^\rho \gamma^{}_\mu \hspace{-0.2cm}&=&\hspace{-0.2cm}
4 g^{\nu\rho} - \left( 4-D \right) \gamma^\nu \gamma^\rho \;,
\nonumber
\\
\gamma^\mu \gamma^\nu \gamma^\rho \gamma^\sigma \gamma^{}_\mu \hspace{-0.2cm}&=&\hspace{-0.2cm}
- 2 \gamma^\sigma \gamma^\rho \gamma^\nu + \left( 4-D \right) \gamma^\nu \gamma^\rho \gamma^\sigma \;.
\end{eqnarray}
Traces of the Dirac matrices which do not contain $\gamma^{}_5$ keep unchanged as
compared with those in the 4-dimensional case; that is,
\begin{eqnarray}
{\rm Tr}\left( \mathbf{1} \right) \hspace{-0.2cm}&=&\hspace{-0.2cm} 4 \;,
\nonumber
\\
{\rm Tr}\left({\rm odd~number~of~} \gamma'{\rm s} \right) \hspace{-0.2cm}&=&\hspace{-0.2cm} 0 \;,
\nonumber
\\
{\rm Tr} \left( \gamma^\mu \gamma^\nu \right) \hspace{-0.2cm}&=&\hspace{-0.2cm} 4 g^{\mu\nu} \;,
\nonumber
\\
{\rm Tr} \left( \gamma^\mu \gamma^\nu \gamma^\rho \gamma^\sigma \right)
\hspace{-0.2cm}&=&\hspace{-0.2cm} 4 \left( g^{\mu\nu} g^{\rho\sigma} - g^{\mu\rho} g^{\nu\sigma}
+ g^{\mu\sigma}g^{\nu\rho} \right) \;,
\end{eqnarray}
where the first relation is just a convention. In $D$ dimensions, the issue of $\gamma^{}_5$
is quite subtle, but we can simply use the ``naive dimensional regularization"~\cite{Buras:1989xd},
namely
\begin{eqnarray}
\left\{\gamma^\mu, \gamma^{}_5 \right\} = 0 \;.
\end{eqnarray}
Although the above relation leads to obvious algebraic inconsistencies~\cite{tHooft:1972tcz,Breitenlohner:1977hr,Breitenlohner:1975hg,Breitenlohner:1976te},
there occurs no trouble concerning the $\gamma^{}_5$ matrix in the present work~\cite{Buras:1989xd}.

In $D$ dimensions, we also need to deal with the dimensions of all the fields and couplings
in the Lagrangian. To keep the action dimensionless, the Lagrangian density should have mass
dimension $D$. For simplicity, we take the QED theory as an example, where the Lagrangian is
given by
\begin{eqnarray}
\mathcal{L}^{}_{\rm QED} = -\frac{1}{4} \left( \partial^{}_\mu A^{}_\nu - \partial^{}_\nu
A^{}_\mu \right)^2 + \overline{\psi} \left({\rm i} \gamma^\mu \partial^{}_\mu - m \right)
\psi - e \overline{\psi} \gamma^\mu \psi A^{}_\mu \;,
\end{eqnarray}
with $A^{}_\mu$ and $\psi$ being the massless gauge field and the massive fermion field,
respectively. The kinetic and mass terms in Eq.~(A6) imply the mass dimensions
\begin{eqnarray}
\left[ m \right] =1 \;, \quad \left[ A^{}_\mu \right] = \frac{D-2}{2} \;,
\quad \left[ \psi \right] = \frac{D-1}{2} \;.
\end{eqnarray}
Then with the help of Eq.~(A7) and the interaction term in Eq.~(A6), one can achieve
$\left[ e \right] = \left( 4-D \right)/2$, where the coupling $e$ is no longer
dimensionless and has a non-integer mass dimension. To keep $e$ dimensionless, it is
conventional to make the replacement
\begin{eqnarray}
e \to \mu^{\frac{4-D}{2}} e \;,
\end{eqnarray}
where $\mu$ is an arbitrary mass-dimension-one parameter. It is also true for the coupling
$g$ in Eq.~(2) that the replacement $g \to \mu^{\left(4-D\right)/2} g$ is made for the
purpose of keeping $g$ dimensionless in $D$ dimensions. Thus in $D$ dimensions, all the
Feynman rules for those vertices involving the couplings $e$ and $g$ should take the
replacements
\begin{eqnarray}
e \to \mu^{\frac{4-D}{2}} e \;, \quad g \to \mu^{\frac{4-D}{2}} g \;.
\end{eqnarray}

\renewcommand{\theequation}{\thesection\arabic{equation}}
\setcounter{equation}{0}
\section{The Passarino-Veltman functions}

Taking account of the on-shell conditions $p^2 = m^2_\mu$, $q^2 = 0$ and $p \cdot
q = \left(m^2_\mu - m^2_e \right)/2$, the physical polarizations
for the external photon $\epsilon \left(q\right) \cdot q = 0$ and
the Passarino-Veltman integrals, we may rewrite the three one-loop Feynman-diagram
amplitudes in Eq.~(5) as follows:
\begin{eqnarray}
{\rm i} \mathcal{M}^{}_{\rm a} \hspace{-0.2cm}&=&\hspace{-0.2cm} \frac{{\rm i} e g^2}{2
\left( 4\pi \right)^2}  \epsilon^\ast_\rho \left( q \right) \overline{u}
\left(p-q\right) \frac{m^{}_e \left( m^{}_e P^{}_{\rm R} + m^{}_\mu P^{}_{\rm L} \right)}
{m^2_\mu - m^2_e} \gamma^\rho u \left(p\right) \sum^{3+n}_{i=1} \mathcal{U}^{}_{ei}
\mathcal{U}^\ast_{\mu i}~ \mu^{ \frac{1}{2} \varepsilon}
\nonumber \\
\hspace{-0.2cm}&&\hspace{-0.2cm} \times \left\{ \left(2-D\right) B^{}_0 \left( p-q
\right) + \left(1-D\right) B^{}_1 \left( p-q \right) + \frac{1}{M^2_W} \left[
\left(D-2\right) B^{}_{00} \left( p-q \right) \right.\right.
\nonumber \\
\hspace{-0.2cm}&&\hspace{-0.2cm} - \left.\left. m^2_e B^{}_{11} \left( p-q \right)
+ A^{}_0 \left( \lambda^{}_i \right) \right] \phantom{\frac{1}{1}} \hspace{-0.35cm}
\right\} \;,
\end{eqnarray}
\begin{eqnarray}
{\rm i} \mathcal{M}^{}_{\rm b} \hspace{-0.2cm}&=&\hspace{-0.2cm} - \frac{{\rm i} e g^2}
{2\left( 4\pi \right)^2}  \epsilon^\ast_\rho \left( q \right) \overline{u}
\left(p-q\right) \frac{m^{}_\mu \left( m^{}_e P^{}_{\rm L} + m^{}_\mu P^{}_{\rm R} \right)}
{m^2_\mu - m^2_e} \gamma^\rho u \left(p\right) \sum^{3+n}_{i=1} \mathcal{U}^{}_{ei}
\mathcal{U}^\ast_{\mu i}~ \mu^{ \frac{1}{2} \varepsilon}
\nonumber \\
\hspace{-0.2cm}&&\hspace{-0.2cm} \times \left\{ \left(2-D\right) B^{}_0 \left(p\right)
+ \left(1-D\right) B^{}_1 \left(p\right) + \frac{1}{M^2_W} \left[ \left(D-2\right)
B^{}_{00} \left(p\right) - m^2_\mu B^{}_{11} \left( p \right) +
A^{}_0 \left( \lambda^{}_i \right) \right] \phantom{\frac{1}{1}} \hspace{-0.35cm} \right\} \;,
\hspace{1cm}
\end{eqnarray}
and
\begin{eqnarray}
{\rm i} \mathcal{M}^{}_{\rm c} \hspace{-0.2cm}&=&\hspace{-0.2cm} - \frac{{\rm i} e g^2}{2\left(
4\pi \right)^2} \epsilon^\ast_\rho \sum^{3+n}_{i=1} \mathcal{U}^{}_{ei}
\mathcal{U}^\ast_{\mu i}~  \overline{u} \left(p-q\right) \mu^{ \frac{1}{2} \varepsilon} \mathcal{M}^\rho \left(p\right) u \left(p\right)
\end{eqnarray}
with
\begin{eqnarray}
\mathcal{M}^\rho \hspace{-0.2cm}&=&\hspace{-0.2cm} 2m^{}_\mu p^\rho P^{}_{\rm R} \left\{
C^{}_0 - C^{}_1 + C^{}_2 - \left(D-1\right) C^{}_{12} + \frac{1}{M^2_W} \left[ B^{}_1
\left(p\right) + B^{}_{11} \left(p\right) + \left(D-2\right) C^{}_{00} + DC^{}_{001}
\right.\right.
\nonumber \\
\hspace{-0.2cm}&&\hspace{-0.2cm} - \left.\left. \hspace{-0.15cm} m^2_e C^{}_{122} \right]
\phantom{\frac{1}{1}} \hspace{-0.35cm} \right\} + 2m^{}_e p^\rho P^{}_{\rm L} \left\{ 2C^{}_0 +
C^{}_1 + D C^{}_2 + \left(D-1\right) \left( C^{}_{12} + C^{}_{22} \right) + \frac{1}{M^2_W}
\left[ - B^{}_1 \left(p\right) -2 C^{}_{00}  \right.\right.
\nonumber \\
\hspace{-0.2cm}&&\hspace{-0.2cm} + \left.\left. \hspace{-0.15cm}  m^2_e C^{}_{22} - D C^{}_{001}
+ 4 C^{}_{002} + 2m^2_e C^{}_{222} + \left( 2m^2_e - m^2_\mu \right) C^{}_{122}  \right]
\phantom{\frac{1}{1}} \hspace{-0.35cm} \right\} - m^{}_e m^{}_\mu \gamma^\rho P^{}_{\rm R}
\left\{\phantom{\frac{1}{1}} \hspace{-0.3cm} 3\left( C^{}_0 + C^{}_2 \right) \right.
\nonumber \\
\hspace{-0.2cm}&&\hspace{-0.2cm} + \left. \frac{1}{M^2_W} \left[ \left(D-4\right) C^{}_{00}
+ m^2_e \left( C^{}_{22} + C^{}_{222} \right) + \left(D+2\right) C^{}_{002} - \left(m^2_\mu
- m^2_e\right) C^{}_{122} \right] \right\}
\nonumber \\
\hspace{-0.2cm}&&\hspace{-0.2cm} - \gamma^\rho P^{}_{\rm L} \left\{\phantom{\frac{1}{1}}
\hspace{-0.35cm} \left( m^2_e + 2 m^2_\mu \right) C^{}_0 + \left(
m^2_\mu - m^2_e \right) C^{}_1 + \left( 2m^2_\mu - m^2_e \right) C^{}_2 - 2\left( D-1 \right)
C^{}_{00} \right.
\nonumber \\
\hspace{-0.2cm}&&\hspace{-0.2cm} + \frac{1}{M^2_W} \left[
\phantom{\frac{1}{1}} \hspace{-0.35cm} \left( m^2_\mu - m^2_e \right)
\left[ 2B^{}_1 \left(p\right) -DC^{}_{001} \right]
+ 2\left( D-1 \right) B^{}_{00} \left(p\right) + 2m^2_\mu B^{}_{11} \left(p\right) \right.
\nonumber \\
\hspace{-0.2cm}&&\hspace{-0.2cm} + \left[ \left(D-2\right) m^2_e - 2m^2_\mu \right] C^{}_{00}
+ \left( m^2_\mu - m^2_e \right)^2 \left( C^{}_{12} +
C^{}_{112} \right) - m^2_e \left( m^2_\mu -2m^2_e \right) C^{}_{22}
\nonumber \\
\hspace{-0.2cm}&&\hspace{-0.2cm} -
\left[ \left(D+2\right) m^2_\mu - 4\left(D+1\right) m^2_e \right] C^{}_{002}
+ \left( m^2_\mu - 5m^2_e \right) \left( m^2_\mu - m^2_e \right) C^{}_{122}
\nonumber \\
\hspace{-0.2cm}&&\hspace{-0.2cm} - \left.\left. m^2_e \left( m^2_\mu - 4m^2_e \right)
C^{}_{222} \phantom{\frac{1}{1}} \hspace{-0.25cm} \right]
\phantom{\frac{1}{1}} \hspace{-0.35cm} \right\} \;,
\end{eqnarray}
where $A^{}_0 \left( \lambda^{}_i \right)$, $B^{}_0 \left( p^\prime \right) \equiv B^{}_0
\left( p^\prime, M^{}_W, \lambda^{}_i \right) $ (for $p^\prime = p$ or $p-q$) and
$C^{}_0 \equiv C^{}_0 \left( q, q-p, M^{}_W, M^{}_W, \lambda^{}_i \right)$ are the
Passarino-Veltman scalar integrals whose generic forms are defined as  \cite{tHooft:1978jhc,Passarino:1978jh,Denner:1991kt}
\begin{eqnarray}
A^{}_0 \left( \lambda \right) \hspace{-0.2cm}&=&\hspace{-0.2cm} \frac{ \left( 2\pi\mu \right)^{ 4-D }}
{{\rm i} \pi^2} \int {\rm d}^D k \frac{1}{k^2 - \lambda^2} \;,
\nonumber \\
B^{}_0 \left( p, \lambda^{}_1, \lambda^{}_2 \right) \hspace{-0.2cm}&=&\hspace{-0.2cm}
\frac{ \left( 2\pi\mu \right)^{ 4-D }}{{\rm i} \pi^2} \int {\rm d}^D k \frac{1}{\left( k^2 - \lambda^2_1 \right) \left[
\left( k+p \right)^2 - \lambda^2_2 \right]} \;,
\nonumber \\
C^{}_0 \left( p^{}_1, p^{}_2, \lambda^{}_1, \lambda^{}_2, \lambda^{}_3 \right)  \hspace{-0.2cm}&=&\hspace{-0.2cm}
\frac{ \left( 2\pi\mu \right)^{ 4-D }}{{\rm i} \pi^2} \int {\rm d}^D k \frac{1}{\left( k^2 - \lambda^2_1
\right) \left[ \left( k+p^{}_1 \right)^2 - \lambda^2_2 \right] \left[ \left( k + p^{}_2 \right)^2
- \lambda^2_3 \right]} \;, \hspace{0.6cm}
\end{eqnarray}
and $B^{}_x \left( p^\prime \right)$ (for $x=1, 00, 11$) and $C^{}_y$
(for $y=1, 2, 00, 11, 12, 22, 001, 002, 112, 122, 222$) are the Passarino-Veltman coefficient
functions for the tension integral decompositions which contain the same parameters as $B^{}_0
\left( p^\prime \right)$ and $C^{}_0$, respectively. Up to the order of
$\mathcal{O}\left(m^2_{\alpha}/M^2_W \right)$ for $\alpha=e$ or $\mu$, the
Passarino-Veltman functions appearing in Eqs.~(B1)---(B4) are explicitly given by
\cite{Patel:2015tea,Patel:2016fam}
\begin{eqnarray}
A^{}_0 \left( \lambda^{}_i \right) = \lambda^2_i \left( \Delta - \ln \frac{M^2_W}{\mu^2} -\ln x^{}_i +1 \right) \; ;
\end{eqnarray}
and
\begin{eqnarray}
B^{}_0 \left( p^\prime \right) \hspace{-0.2cm}&\simeq&\hspace{-0.2cm} 1 + \Delta - \frac{x^{}_i}{x^{}_i-1}
\ln x^{}_i + \frac{ x^2_i-1-2x^{}_i\ln x^{}_i}{ 2\left( x^{}_i-1 \right)^3} \cdot \frac{m^2_\alpha}{M^2_W} - \ln \frac{M^2_W}{\mu^2} \;,
\nonumber \\
B^{}_1 \left( p^\prime \right) \hspace{-0.2cm}&\simeq&\hspace{-0.2cm} -\frac{1}{2} \Delta + \frac{-x^2_i +
4x^{}_i - 3 + 2x^{}_i \left(x^{}_i - 2\right) \ln x^{}_i}{4\left(x^{}_i-1\right)^2} - \frac{x^3_i -
6x^2_i + 3x^{}_i + 2 + 6x^{}_i \ln x^{}_i }{6\left(x^{}_i-1\right)^4} \cdot \frac{m^2_\alpha}{M^2_W}
\nonumber \\
\hspace{-0.2cm}&&\hspace{-0.2cm} + \frac{1}{2} \ln \frac{M^2_W}{\mu^2} \;,
\end{eqnarray}
\begin{eqnarray}
B^{}_{00} \left( p^\prime \right) \hspace{-0.2cm}&\simeq&\hspace{-0.2cm}  \frac{3M^2_W\left( x^{}_i + 1
\right)- m^2_\alpha}{12} \Delta + \left[ \frac{1}{12} \ln \frac{M^2_W}{\mu^2} + \frac{-5x^3_i + 27x^2_i
-27x^{}_i + 5 + 6x^2_i \left(x^{}_i - 3\right) \ln x^{}_i}{72\left(x^{}_i-1\right)^3} \right] m^2_\alpha
\nonumber \\
\hspace{-0.2cm}&&\hspace{-0.2cm} + \frac{3\left(x^2_i-1\right) - 2x^2_i \ln x^{}_i }{8\left(x^{}_i-1
\right)} M^2_W - \frac{1}{4} M^2_W \left( x^{}_i + 1 \right) \ln \frac{M^2_W}{\mu^2} \;,
\nonumber \\
B^{}_{11} \left( p^\prime \right) \hspace{-0.2cm}&\simeq&\hspace{-0.2cm} \frac{1}{3} \Delta + \frac{2x^3_i
- 9x^2_i + 18x^{}_i -11 - 6x^{}_i \left(x^2_i -3x^{}_i + 3 \right) \ln x^{}_i }{18\left(x^{}_i-1
\right)^3}
\nonumber \\
\hspace{-0.2cm}&&\hspace{-0.2cm} + \frac{x^4_i - 6x^3_i + 18x^2_i - 10x^{}_i -3 - 12x^{}_i \ln x^{}_i }{12\left(x^{}_i-1\right)^5} \cdot \frac{m^2_\alpha}{M^2_W}
- \frac{1}{3} \ln \frac{M^2_W}{\mu^2} \; ;
\end{eqnarray}
and
\begin{eqnarray}
C^{}_0 \hspace{-0.2cm}&\simeq&\hspace{-0.2cm} \frac{x^{}_i-1-x^{}_i \ln x^{}_i }{M^2_W\left(x^{}_i-1\right)^2}
+\left(m^2_e + m^2_\mu \right) \frac{5x^2_i - 4x^{}_i -1 - 2x^{}_i \left(x^{}_i+2\right) \ln x^{}_i }
{4M^4_W\left(x^{}_i-1\right)^2} \;,
\nonumber \\
C^{}_1 \hspace{-0.2cm}&\simeq&\hspace{-0.2cm} - \frac{3x^2_i - 4x^{}_i + 1 - 2x^2_i \ln x^{}_i}{4M^2_W\left(
x^{}_i-1\right)^3} + \left( m^2_e + 2m^2_\mu \right) \frac{-17x^3_i + 9x^2_i + 9x^{}_i -1 + 6x^2_i \left(
x^{}_i + 3\right) \ln x^{}_i}{36M^4_W \left(x^{}_i -1 \right)^5} \;,
\nonumber \\
C^{}_2 \hspace{-0.2cm}&\simeq&\hspace{-0.2cm} \frac{x^2_i - 1 -2x^{}_i \ln x^{}_i}{2M^2_W\left(x^{}_i-1\right)^3}
+ \left( m^2_e + m^2_\mu \right) \frac{x^3_i + 9x^2_i - 9x^{}_i - 1 - 6x^{}_i \left(x^{}_i + 1\right) \ln
x^{}_i}{6M^4_W\left(x^{}_i-1\right)^5} \;,
\end{eqnarray}
\begin{eqnarray}
C^{}_{00} \hspace{-0.2cm}&\simeq&\hspace{-0.2cm} \frac{1}{4} \Delta + \frac{3x^2_i - 4x^{}_i + 1 - 2x^2_i \ln
x^{}_i}{8\left(x^{}_i-1\right)^2} + \left( m^2_e + m^2_\mu \right) \frac{2x^3_i + 3x^2_i - 6x^{}_i + 1 -
6x^2_i \ln x^{}_i}{24M^2_W\left(x^{}_i-1\right)^4} - \frac{1}{4} \ln \frac{M^2_W}{\mu^2} \;,
\nonumber \\
C^{}_{12} \hspace{-0.2cm}&\simeq&\hspace{-0.2cm} -\frac{2x^3_i + 3x^2_i - 6x^{}_i + 1 - 6x^2_i \ln x^{}_i }
{12M^2_W\left(x^{}_i-1\right)^4}
\nonumber \\
\hspace{-0.2cm}&&\hspace{-0.2cm}
- \left( m^2_e + 2m^2_\mu \right) \frac{3x^4_i + 44x^3_i - 36x^2_i -
12x^{}_i + 1 - 12x^2_i \left(2x^{}_i + 3 \right) \ln x^{}_i }{72M^4_W\left(x^{}_i-1\right)^6} \;,
\nonumber \\
C^{}_{22} \hspace{-0.2cm}&\simeq&\hspace{-0.2cm} -\frac{x^3_i - 6x^2_i + 3x^{}_i + 2 + 6x^{}_i \ln x^{}_i
}{6M^2_W\left(x^{}_i-1\right)^4}
\nonumber \\
\hspace{-0.2cm}&&\hspace{-0.2cm}
- \left(m^2_e +m^2_\mu\right) \frac{x^4_i - 12x^3_i - 36x^2_i +
44x^{}_i + 3 + 12x^{}_i \left(3x^{}_i + 2 \right) \ln x^{}_i }{24M^4_W\left(x^{}_i-1\right)^6} \;,
\end{eqnarray}
\begin{eqnarray}
C^{}_{001} \hspace{-0.2cm}&\simeq&\hspace{-0.2cm} -\frac{1}{12} \Delta - \frac{11x^3_i - 18x^2_i + 9x^{}_i
- 2 - 6x^3_i \ln x^{}_i}{72\left(x^{}_i-1\right)^3}
\nonumber \\
\hspace{-0.2cm}&&\hspace{-0.2cm}
- \left( m^2_e + 2m^2_\mu \right) \frac{3x^4_i
+ 10x^3_i - 18x^2_i + 6x^{}_i - 1 - 12x^3_i \ln x^{}_i }{144M^2_W\left(x^{}_i-1\right)^5}
+ \frac{1}{12} \ln \frac{M^2_W}{\mu^2} \;,
\nonumber \\
C^{}_{002} \hspace{-0.2cm}&\simeq&\hspace{-0.2cm} -\frac{1}{12} \Delta - \frac{5x^3_i - 27x^2_i + 27x^{}_i
- 5 - 6x^2_i \left(x^{}_i-3\right) \ln x^{}_i}{72\left(x^{}_i-1\right)^3}
\nonumber \\
\hspace{-0.2cm}&&\hspace{-0.2cm}
- \left( m^2_e + m^2_\mu \right) \frac{x^4_i - 8x^3_i + 8x^{}_i - 1 + 12x^2_i \ln x^{}_i }{48M^2_W\left(x^{}_i-1\right)^5} + \frac{1}{12} \ln \frac{M^2_W}{\mu^2} \;,
\nonumber \\
C^{}_{112} \hspace{-0.2cm}&\simeq&\hspace{-0.2cm} \frac{3x^4_i + 10x^3_i - 18x^2_i + 6x^{}_i - 1 -
12x^3_i \ln x^{}_i}{36M^2_W\left(x^{}_i-1\right)^5}
\nonumber \\
\hspace{-0.2cm}&&\hspace{-0.2cm} + \left(m^2_e + 3m^2_\mu\right) \frac{6x^5_i + 125x^4_i - 80x^3_i
- 60x^2_i + 10x^{}_i - 1 - 60 x^3_i \left( x^{}_i + 2 \right) \ln x^{}_i}{360M^4_W\left(x^{}_i-1
\right)^7} \;,
\nonumber \\
C^{}_{122} \hspace{-0.2cm}&\simeq&\hspace{-0.2cm} \frac{x^4_i - 8x^3_i + 8x^{}_i - 1 + 12x^2_i \ln
x^{}_i }{24M^2_W\left(x^{}_i-1\right)^5}
\nonumber \\
\hspace{-0.2cm}&&\hspace{-0.2cm} + \left( m^2_e + 2m^2_\mu \right) \frac{x^5_i - 15x^4_i - 80x^3_i
+ 80x^2_i + 15x^{}_i -1 + 60x^2_i \left(x^{}_i + 1 \right) \ln x^{}_i }{120M^4_W\left(x^{}_i-1
\right)^7} \;,
\nonumber \\
C^{}_{222} \hspace{-0.2cm}&\simeq&\hspace{-0.2cm} \frac{x^4_i - 6x^3_i + 18x^2_i -10x^{}_i - 3 -
12x^{}_i \ln x^{}_i }{12M^2_W\left(x^{}_i-1\right)^5}
\nonumber \\
\hspace{-0.2cm}&&\hspace{-0.2cm} + \left( m^2_e + m^2_\mu \right) \frac{x^5_i - 10x^4_i + 60x^3_i
+ 80x^2_i -125 x^{}_i - 6 - 60x^{}_i \left(2x^{}_i + 1 \right) \ln x^{}_i }{60M^4_W\left(x^{}_i-1
\right)^7} \; ,
\end{eqnarray}
where $\alpha = e$ (or $\mu$) when $p^\prime = p-q$ (or $p$), $x^{}_i = \lambda^2_i / M^2_W$ and
$\Delta \equiv 2/\varepsilon - \gamma^{}_{\rm E} + \ln \left(4\pi\right)$ with $\gamma^{}_{\rm E}$
being Euler's constant.

\renewcommand{\theequation}{\thesection\arabic{equation}}
\setcounter{equation}{0}
\section{The Jarlskog-like invariants}

In the $(3+3)$ active-sterile neutrino mixing scheme, the $6\times6$ unitary matrix
$\mathcal{U}$ can be expressed as \cite{Xing:2011ur}
\begin{eqnarray}
\mathcal{U} = \left(\begin{matrix} I & 0 \\ 0 & U^\prime_0 \end{matrix} \right)
\left(\begin{matrix} A & R \\ S & B \end{matrix} \right) \left(\begin{matrix} U^{}_0 & 0
\\ 0 & I \end{matrix} \right) \;,
\end{eqnarray}
where $I$ denotes the $3\times3$ identity matrix, $U^{}_0$ and $U^\prime_0$ are unitary
matrices responsible respectively for flavor mixing in the active sector and that in the
sterile sector, and $A$, $B$, $R$ and $S$ are the $3\times3$ matrices describing the interplay
between the two sectors. Then an Euler-like parametrization of $\mathcal{U}$ is
\begin{eqnarray}
\left(\begin{matrix} U^{}_0 & 0 \\ 0 & I \end{matrix} \right) \hspace{-0.2cm}&=&\hspace{-0.2cm}
O^{}_{23} O^{}_{13} O^{}_{12} \;,
\nonumber \\
\left(\begin{matrix} I & 0 \\ 0 & U^\prime_0 \end{matrix} \right) \hspace{-0.2cm}&=&\hspace{-0.2cm}
O^{}_{56} O^{}_{46} O^{}_{45} \;,
\nonumber \\
\left(\begin{matrix} A & R \\ S & B \end{matrix} \right) \hspace{-0.2cm}&=&\hspace{-0.2cm} O^{}_{36}
O^{}_{26} O^{}_{16} O^{}_{35} O^{}_{25} O^{}_{15} O^{}_{34} O^{}_{24} O^{}_{14} \;,
\end{eqnarray}
where $O^{}_{ij}$ (for $1 \leq i < j \leq 6$) are the two-dimensional $6 \times 6$
rotation matrices in the complex plane \cite{Xing:2007zj,Xing:2011ur}.
The elements of $O^{}_{ij}$ satisfy $O^{}_{ij}
\left( i,i\right) = O^{}_{ij} \left( j,j\right) = \cos \theta^{}_{ij} $, $O^{}_{ij}
\left( i,j\right) = - O^{\ast}_{ij} \left( j,i\right)=\sin \theta^{}_{ij} \exp
\left( -{\rm i} \delta^{}_{ij} \right) $, $O^{}_{ij} \left(n,n\right) = 1$ (for $n\neq i, j$),
and all the other elements are vanishing. So there are totally fifteen mixing angles
$\theta^{}_{ij}$ and fifteen CP-violating phases $\delta^{}_{ij}$ in the parametrization
of $\mathcal{U}$. The PMNS matrix $U$ is actually a product of $A$ and $U^{}_0$; namely,
$U \equiv A U^{}_0$, where $A$ measures the departure of $U$ from $U^{}_0$.
The unitarity of $\cal U$ assures $UU^\dagger + RR^\dagger = AA^\dagger + RR^\dagger = I$,
and the exact canonical seesaw formula is given by
\begin{eqnarray}
U D^{}_\nu U^T + R D^{}_N R^T = 0 \;,
\end{eqnarray}
where $D^{}_\nu \equiv {\rm Diag} \{ m^{}_1, m^{}_2, m^{}_3 \}$ and $D^{}_N \equiv {\rm Diag}
\{ M^{}_1, M^{}_2, M^{}_3\}$ are defined to be the diagonal active and sterile neutrino
mass matrices, respectively. To calculate the Jarlskog-like invariants
${\cal J}^{ij}_{\alpha\beta}$ defined in Eq.~(4), let us write out the expressions
of $U^{}_0$ and $A$ \cite{Xing:2019vks}:
\begin{eqnarray}
U^{}_0 = \left( \begin{matrix} c^{}_{12} c^{}_{13} & \hat{s}^*_{12}
c^{}_{13} & \hat{s}^*_{13} \cr \vspace{-0.45cm} \cr
-\hat{s}^{}_{12} c^{}_{23} -
c^{}_{12} \hat{s}^{}_{13} \hat{s}^*_{23} & c^{}_{12} c^{}_{23} -
\hat{s}^*_{12} \hat{s}^{}_{13} \hat{s}^*_{23} & c^{}_{13}
\hat{s}^*_{23} \cr \vspace{-0.45cm} \cr
\hat{s}^{}_{12} \hat{s}^{}_{23} - c^{}_{12}
\hat{s}^{}_{13} c^{}_{23} & -c^{}_{12} \hat{s}^{}_{23} -
\hat{s}^*_{12} \hat{s}^{}_{13} c^{}_{23} & c^{}_{13} c^{}_{23}
\cr \end{matrix} \right) \; ,
\end{eqnarray}
and
\begin{eqnarray}
A = \left( \begin{matrix} c^{}_{14} c^{}_{15} c^{}_{16} & 0 & 0 \cr \vspace{-0.35cm} \cr
\begin{array}{l} -c^{}_{14} c^{}_{15} \hat{s}^{}_{16} \hat{s}^*_{26} -
c^{}_{14} \hat{s}^{}_{15} \hat{s}^*_{25} c^{}_{26} \\
-\hat{s}^{}_{14} \hat{s}^*_{24} c^{}_{25} c^{}_{26} \end{array} &
c^{}_{24} c^{}_{25} c^{}_{26} & 0 \cr \vspace{-0.35cm} \cr
\begin{array}{l} -c^{}_{14} c^{}_{15} \hat{s}^{}_{16} c^{}_{26} \hat{s}^*_{36}
+ c^{}_{14} \hat{s}^{}_{15} \hat{s}^*_{25} \hat{s}^{}_{26} \hat{s}^*_{36} \\
- c^{}_{14} \hat{s}^{}_{15} c^{}_{25} \hat{s}^*_{35} c^{}_{36} +
\hat{s}^{}_{14} \hat{s}^*_{24} c^{}_{25} \hat{s}^{}_{26}
\hat{s}^*_{36} \\
+ \hat{s}^{}_{14} \hat{s}^*_{24} \hat{s}^{}_{25} \hat{s}^*_{35}
c^{}_{36} - \hat{s}^{}_{14} c^{}_{24} \hat{s}^*_{34} c^{}_{35}
c^{}_{36} \end{array} &
\begin{array}{l} -c^{}_{24} c^{}_{25} \hat{s}^{}_{26} \hat{s}^*_{36} -
c^{}_{24} \hat{s}^{}_{25} \hat{s}^*_{35} c^{}_{36} \\
-\hat{s}^{}_{24} \hat{s}^*_{34} c^{}_{35} c^{}_{36} \end{array} &
c^{}_{34} c^{}_{35} c^{}_{36} \cr \end{matrix} \right) \; ,
\end{eqnarray}
where $\hat{s}^{}_{ij} \equiv e^{{\rm i}\delta^{}_{ij}} \sin\theta^{}_{ij}$.
Since the active-sterile neutrino mixing angles $\theta^{}_{ij}$ (for
$i = 1,2,3$ and $j = 4,5,6$) are expected to at most of ${\cal O}(0.1)$,
one may make some reasonable analytical approximations to simplify the
elements of $A$ up to ${\cal O}(10^{-2})$. Then we obtain the
results of nine Jarlskog-like invarants as follows:
\begin{eqnarray}
\mathcal{J}^{12}_{e\mu} \hspace{-0.2cm}&\simeq&\hspace{-0.2cm} \left(1-\mathcal{X}^\prime\right)
\mathcal{J}^{}_0 + c^{}_{12} s^{}_{12} c^3_{13} c^{}_{23} {\rm Im} X^{}_1\;,
\nonumber \\
\mathcal{J}^{23}_{e\mu} \hspace{-0.2cm}&\simeq&\hspace{-0.2cm} \left(1-\mathcal{X}^\prime\right)
\mathcal{J}^{}_0 - s^{}_{12} c^{}_{13} s^{}_{13} \left( c^{}_{12} s^{}_{13} c^{}_{23}
{\rm Im} X^{}_1 - s^{}_{12} s^{}_{23} {\rm Im} X^{}_2 \right) \;,
\nonumber \\
\mathcal{J}^{31}_{e\mu} \hspace{-0.2cm}&\simeq&\hspace{-0.2cm} \left(1-\mathcal{X}^\prime\right)
\mathcal{J}^{}_0 - c^{}_{12} c^{}_{13} s^{}_{13} \left( s^{}_{12} s^{}_{13} c^{}_{23}
{\rm Im} X^{}_1 + c^{}_{12} s^{}_{23} {\rm Im} X^{}_2 \right) \;, \hspace{0.5cm}
\end{eqnarray}
and
\begin{eqnarray}
\mathcal{J}^{12}_{\tau e} \hspace{-0.2cm}&\simeq&\hspace{-0.2cm} \left(1-\mathcal{Y}^\prime\right)
\mathcal{J}^{}_0 + c^{}_{12} s^{}_{12} c^3_{13} s^{}_{23} {\rm Im} Y^{}_1 - c^{}_{12}
s^{}_{12} c^2_{13} s^{}_{13} \left(  c^2_{23} {\rm Im} Z^{}_2 + s^2_{23} {\rm Im}
Z^{}_3 \right) \;,
\nonumber \\
\mathcal{J}^{23}_{\tau e} \hspace{-0.2cm}&\simeq&\hspace{-0.2cm} \left(1-\mathcal{Y}^\prime\right)
\mathcal{J}^{}_0 - s^{}_{12} c^{}_{13} s^{}_{13} \left( c^{}_{12} s^{}_{13} s^{}_{23}
{\rm Im} Y^{}_1 + s^{}_{12} c^{}_{23} {\rm Im} Y^{}_2 \right) - c^{}_{12} s^{}_{12}
c^2_{13} s^{}_{13} \left(  c^2_{23} {\rm Im} Z^{}_2 + s^2_{23} {\rm Im} Z^{}_3 \right) \;,
\nonumber \\
\mathcal{J}^{31}_{\tau e} \hspace{-0.2cm}&\simeq&\hspace{-0.2cm} \left(1-\mathcal{Y}^\prime\right)
\mathcal{J}^{}_0 - c^{}_{12} c^{}_{13} s^{}_{13} \left( s^{}_{12} s^{}_{13} s^{}_{23}
{\rm Im} Y^{}_1 - c^{}_{12} c^{}_{23} {\rm Im} Y^{}_2 \right) - c^{}_{12} s^{}_{12}
c^2_{13} s^{}_{13} \left(  c^2_{23} {\rm Im} Z^{}_2 + s^2_{23} {\rm Im} Z^{}_3 \right) \;, \hspace{0.5cm}
\end{eqnarray}
as well as
\begin{eqnarray}
\mathcal{J}^{12}_{\mu \tau} \hspace{-0.2cm}&\simeq&\hspace{-0.2cm} \left(1-\mathcal{Z}^\prime\right)
\mathcal{J}^{}_0 + c^{}_{12} s^{}_{12} c^{}_{13} \left[ c^{}_{23} \left( s^2_{23} -
s^2_{13} \right) {\rm Im} X^{}_1  + s^{}_{23} \left( c^2_{23} - s^2_{13} \right) {\rm Im}
Y^{}_1 \right]
\nonumber \\
\hspace{-0.2cm}&&\hspace{-0.2cm} - \left( c^2_{12} - s^2_{12} \right) c^{}_{13} s^{}_{13}
c^{}_{23} s^{}_{23} \left( c^{}_{23} {\rm Im} X^{}_2 - s^{}_{23} {\rm Im} Y^{}_2  \right)
+ c^{}_{12} s^{}_{12} c^{}_{13} s^2_{13} c^{}_{23} s^{}_{23} \left( s^{}_{23} {\rm Im}
X^{}_3 + c^{}_{23} {\rm Im} Y^{}_3 \right)
\nonumber \\
\hspace{-0.2cm}&&\hspace{-0.2cm} - \left( c^2_{12} - s^2_{12} \right) s^2_{13} c^{}_{23}
s^{}_{23} {\rm Im} Z^{}_1 - c^{}_{12} s^{}_{12} s^{}_{13} c^2_{23} {\rm Im} Z^{}_2 +
c^{}_{12} s^{}_{12} s^3_{13} s^2_{23} {\rm Im} Z^{}_3 \;,
\nonumber \\
\mathcal{J}^{23}_{\mu \tau} \hspace{-0.2cm}&\simeq&\hspace{-0.2cm} \left(1-\mathcal{Z}^\prime\right)
\mathcal{J}^{}_0 + c^{}_{12} c^{}_{13} \left[s^{}_{12} c^{}_{23} \left( s^2_{23} - s^2_{13}
\right) {\rm Im} X^{}_1  + s^{}_{12} s^{}_{23} \left( c^2_{23} - s^2_{13} \right) {\rm Im}
Y^{}_1 + c^{}_{12} c^{}_{13} s^{}_{23} c^{}_{23} {\rm Im} Z^{}_1 \right]
\nonumber
\\
\hspace{-0.2cm}&&\hspace{-0.2cm} - \left( c^2_{12} - s^2_{12} \right) c^{}_{13} s^{}_{13}
c^{}_{23} s^{}_{23} \left( c^{}_{23} {\rm Im} X^{}_2 - s^{}_{23} {\rm Im} Y^{}_2  \right)
+ c^{}_{12} s^{}_{12} c^{}_{13} s^2_{13} c^{}_{23} s^{}_{23} \left( s^{}_{23} {\rm Im}
X^{}_3 + c^{}_{23} {\rm Im} Y^{}_3 \right)
\nonumber \\
\hspace{-0.2cm}&&\hspace{-0.2cm}  - c^{}_{12} s^{}_{12} c^2_{13} s^{}_{13} s^2_{23}
{\rm Im} Z^{}_3 \;,
\nonumber \\
\mathcal{J}^{31}_{\mu \tau} \hspace{-0.2cm}&\simeq&\hspace{-0.2cm} \left(1-\mathcal{Z}^\prime\right)
\mathcal{J}^{}_0 + s^{}_{12} c^{}_{13} \left[c^{}_{12} c^{}_{23} \left( s^2_{23} - s^2_{13}
\right) {\rm Im} X^{}_1  + c^{}_{12} s^{}_{23} \left( c^2_{23} - s^2_{13} \right)
{\rm Im} Y^{}_1 - s^{}_{12} c^{}_{13} s^{}_{23} c^{}_{23} {\rm Im} Z^{}_1 \right] \hspace{0.5cm}
\nonumber \\
\hspace{-0.2cm}&&\hspace{-0.2cm} - \left( c^2_{12} - s^2_{12} \right) c^{}_{13} s^{}_{13}
c^{}_{23} s^{}_{23} \left( c^{}_{23} {\rm Im} X^{}_2 - s^{}_{23} {\rm Im} Y^{}_2  \right)
+ c^{}_{12} s^{}_{12} c^{}_{13} s^2_{13} c^{}_{23} s^{}_{23} \left( s^{}_{23} {\rm Im} X^{}_3
+ c^{}_{23} {\rm Im} Y^{}_3 \right)
\nonumber \\
\hspace{-0.2cm}&&\hspace{-0.2cm}  - c^{}_{12} s^{}_{12} c^2_{13} s^{}_{13} s^2_{23}
{\rm Im} Z^{}_3 \;,
\end{eqnarray}
where $\mathcal{J}^{}_0 = c^{}_{12} s^{}_{12} c^2_{13} s^{}_{13} c^{}_{23} s^{}_{23}
\sin \delta$ with $\delta \equiv \delta^{}_{13} - \delta^{}_{12} -\delta^{}_{23}$, and
\begin{eqnarray}
X^{}_1 \hspace{-0.2cm}&=&\hspace{-0.2cm} \mathcal{X} e^{-{\rm i}\delta^{}_{12}}, \quad
X^{}_2 = \mathcal{X} e^{-{\rm i} \left( \delta^{}_{13} - \delta^{}_{23}  \right)} ,\quad
X^{}_3 = \mathcal{X}  e^{{\rm i} \left( \delta^{}_{12} - 2\delta^{}_{13} + 2 \delta^{}_{23}
\right)} \;,
\nonumber \\
Y^{}_1 \hspace{-0.2cm}&=&\hspace{-0.2cm} \mathcal{Y} e^{-{\rm i} \left( \delta^{}_{12} +
\delta^{}_{23} \right) } , \quad Y^{}_2 = \mathcal{Y} e^{-{\rm i}\delta^{}_{13}} ,\quad
Y^{}_3 = \mathcal{Y}  e^{{\rm i} \left( \delta^{}_{12} - 2\delta^{}_{13} + \delta^{}_{23}
\right)} \;,
\nonumber \\
Z^{}_1 \hspace{-0.2cm}&=&\hspace{-0.2cm} \mathcal{Z} e^{-{\rm i} \delta^{}_{23}} , \quad
Z^{}_2 = \mathcal{Z} e^{{\rm i} \left( \delta^{}_{12} - \delta^{}_{13} \right) } , \quad
Z^{}_3 = \mathcal{Z} e^{-{\rm i} \left( \delta^{}_{12} - \delta^{}_{13} + 2 \delta^{}_{23}
\right)} \;, \hspace{0.5cm}
\end{eqnarray}
together with
\begin{eqnarray}
\mathcal{X} \hspace{-0.2cm}&\equiv&\hspace{-0.2cm} \hat{s}^{}_{14} \hat{s}^\ast_{24} +
\hat{s}^{}_{15} \hat{s}^\ast_{25} + \hat{s}^{}_{16} \hat{s}^\ast_{26} , \quad
\mathcal{X}^\prime \equiv s^2_{14} + s^2_{15} + s^2_{16} + s^2_{24} + s^2_{25} + s^2_{26} \;,
\nonumber \\
\mathcal{Y} \hspace{-0.2cm}&\equiv&\hspace{-0.2cm} \hat{s}^{}_{14} \hat{s}^\ast_{34} +
\hat{s}^{}_{15} \hat{s}^\ast_{35} + \hat{s}^{}_{16} \hat{s}^\ast_{36} , \quad
\mathcal{Y}^\prime \equiv s^2_{14} + s^2_{15} + s^2_{16} + s^2_{34} + s^2_{35} + s^2_{36} \;,
\nonumber \\
\mathcal{Z} \hspace{-0.2cm}&\equiv&\hspace{-0.2cm} \hat{s}^{}_{24} \hat{s}^\ast_{34} +
\hat{s}^{}_{25} \hat{s}^\ast_{35} + \hat{s}^{}_{26} \hat{s}^\ast_{36} , \quad
\mathcal{Z}^\prime \equiv s^2_{24} + s^2_{25} + s^2_{26} + s^2_{34} + s^2_{35} + s^2_{36} \;.
\hspace{0.5cm}
\end{eqnarray}
If the smallness of $\theta^{}_{13}$ is considered and the terms of $\mathcal{O}
\left(s^{}_{13} |\mathcal{X}^{(\prime)}| \right)$, $\mathcal{O} \left(s^{}_{13}
|\mathcal{Y}^{(\prime)}| \right)$
and $\mathcal{O} \left(s^{}_{13} |\mathcal{Z}^{(\prime)}| \right)$
together with those higher-order terms are omitted in Eqs.~(C6)---(C8), then the results
in Ref.~\cite{Xing:2011ur} can be reproduced and one will be left with
$\mathcal{J}^{23}_{e\mu} \simeq \mathcal{J}^{31}_{e\mu} \simeq \mathcal{J}^{23}_{\tau e}
\simeq \mathcal{J}^{31}_{\tau e} \simeq \mathcal{J}^{}_0$.


\newpage

\begin{thebibliography}{99}
\bibitem{Zyla:2020zbs}
P.~A.~Zyla \textit{et al.} [Particle Data Group],
PTEP \textbf{2020} (2020) no.8, 083C01.

\bibitem{Xing:2019vks}
Z.~z.~Xing,
Phys. Rept. \textbf{854} (2020) 1
[arXiv:1909.09610 [hep-ph]].

\bibitem{Pontecorvo:1957cp}
  B.~Pontecorvo,
  Sov.\ Phys.\ JETP {\bf 6} (1957) 429
   [Zh.\ Eksp.\ Teor.\ Fiz.\  {\bf 33} (1957) 549].

\bibitem{Maki:1962mu}
  Z.~Maki, M.~Nakagawa and S.~Sakata,
  Prog.\ Theor.\ Phys.\  {\bf 28} (1962) 870.

\bibitem{Pontecorvo:1967fh}
  B.~Pontecorvo,
  Sov.\ Phys.\ JETP {\bf 26} (1968) 984
   [Zh.\ Eksp.\ Teor.\ Fiz.\  {\bf 53} (1967) 1717].

\bibitem{Adey:2018zwh}
  D.~Adey \textit{et al.} [Daya Bay Collaboration],
  Phys. Rev. Lett. \textbf{121} (2018) 241805
  [arXiv:1809.02261 [hep-ex]].

\bibitem{Cabibbo:1963yz}
  N.~Cabibbo,
  Phys. Rev. Lett. \textbf{10} (1963) 531.

\bibitem{Kobayashi:1973fv}
  M.~Kobayashi and T.~Maskawa,
  Prog. Theor. Phys. \textbf{49} (1973) 652.

\bibitem{Abe:2019vii}
  K.~Abe {\it et al.} [T2K Collaboration],
  Nature {\bf 580} (2020) 339
  [arXiv:1910.03887 [hep-ex]].

\bibitem{Fritzsch:1999ee}
  H.~Fritzsch and Z.~z.~Xing,
  Prog. Part. Nucl. Phys. \textbf{45} (2000) 1
  [arXiv:hep-ph/9912358 [hep-ph]].

\bibitem{AguilarSaavedra:2000vr}
  J.~Aguilar-Saavedra and G.~Branco,
  Phys. Rev. D \textbf{62} (2000) 096009
  [arXiv:hep-ph/0007025 [hep-ph]].

\bibitem{Xing:2005gk}
  Z.~z.~Xing and H.~Zhang,
  Phys. Lett. B \textbf{618} (2005) 131
  [arXiv:hep-ph/0503118 [hep-ph]].

\bibitem{Luo:2011mm}
S.~Luo,
Phys. Rev. D \textbf{85} (2012), 013006
[arXiv:1109.4260 [hep-ph]].

\bibitem{Xing:2015wzz}
Z.~z.~Xing and J.~y.~Zhu,
Nucl. Phys. B \textbf{908} (2016), 302-317
[arXiv:1511.00450 [hep-ph]].

\bibitem{Esteban:2016qun}
I.~Esteban, M.~C.~Gonzalez-Garcia, M.~Maltoni, I.~Martinez-Soler and T.~Schwetz,
JHEP \textbf{01} (2017), 087
[arXiv:1611.01514 [hep-ph]].

\bibitem{Zhu:2018dvj}
  J.~Y.~Zhu,
  Phys. Rev. D \textbf{99} (2019) 033003
  [arXiv:1810.04426 [hep-ph]].

\bibitem{Xing:2019tsn}
Z.~Z.~Xing and D.~Zhang,
Phys. Lett. B \textbf{803} (2020), 135302
[arXiv:1911.03292 [hep-ph]].

\bibitem{Ellis:2020ehi}
  S.~A.~R.~Ellis, K.~J.~Kelly and S.~W.~Li,
  [arXiv:2004.13719 [hep-ph]].

\bibitem{Ellis:2020hus}
S.~A.~R.~Ellis, K.~J.~Kelly and S.~W.~Li,
[arXiv:2008.01088 [hep-ph]].

\bibitem{Minkowski:1977sc}
  P.~Minkowski,
  Phys.\ Lett.\  {\bf 67B} (1977) 421.

\bibitem{Yanagida:1979as}
  T.~Yanagida,
  Conf.\ Proc.\ C {\bf 7902131} (1979) 95.

\bibitem{GellMann:1980vs}
  M.~Gell-Mann, P.~Ramond and R.~Slansky,
  Conf.\ Proc.\ C {\bf 790927} (1979) 315
  [arXiv:1306.4669 [hep-th]].

\bibitem{Glashow:1979nm}
  S.~L.~Glashow,
  NATO Sci.\ Ser.\ B {\bf 61} (1980) 687.

\bibitem{Mohapatra:1979ia}
  R.~N.~Mohapatra and G.~Senjanovic,
  Phys.\ Rev.\ Lett.\  {\bf 44} (1980) 912.

\bibitem{Xing:2007zj}
  Z.~z.~Xing,
  Phys. Lett. B \textbf{660} (2008) 515
  [arXiv:0709.2220 [hep-ph]].

\bibitem{Xing:2011ur}
Z.~z.~Xing,
Phys. Rev. D \textbf{85} (2012), 013008
[arXiv:1110.0083 [hep-ph]].

\bibitem{Lindner:2016bgg}
  M.~Lindner, M.~Platscher and F.~S.~Queiroz,
  Phys. Rept. \textbf{731} (2018), 1-82
  [arXiv:1610.06587 [hep-ph]].

\bibitem{Ilakovac:1994kj}
  A.~Ilakovac and A.~Pilaftsis,
  Nucl. Phys. B \textbf{437} (1995), 491
  [arXiv:hep-ph/9403398 [hep-ph]].

\bibitem{Alonso:2012ji}
  R.~Alonso, M.~Dhen, M.~B.~Gavela and T.~Hambye,
  JHEP \textbf{01} (2013), 118
  [arXiv:1209.2679 [hep-ph]].

\bibitem{Petcov:1976ff}
S.~T.~Petcov,
Sov. J. Nucl. Phys. \textbf{25} (1977), 340
JINR-E2-10176.

\bibitem{Bilenky:1977du}
S.~M.~Bilenky, S.~T.~Petcov and B.~Pontecorvo,
Phys. Lett. B \textbf{67} (1977), 309.

\bibitem{Cheng:1976uq}
T.~P.~Cheng and L.~F.~Li,
Phys. Rev. Lett. \textbf{38} (1977), 381.

\bibitem{Marciano:1977wx}
W.~J.~Marciano and A.~I.~Sanda,
Phys. Lett. B \textbf{67} (1977), 303-305.

\bibitem{Lee:1977qz}
B.~W.~Lee, S.~Pakvasa, R.~E.~Shrock and H.~Sugawara,
Phys. Rev. Lett. \textbf{38} (1977), 937.

\bibitem{Lee:1977tib}
B.~W.~Lee and R.~E.~Shrock,
Phys. Rev. D \textbf{16} (1977), 1444.

\bibitem{Antusch:2006vwa}
  S.~Antusch, C.~Biggio, E.~Fernandez-Martinez, M.~B.~Gavela and J.~Lopez-Pavon,
  JHEP \textbf{10} (2006), 084
  [arXiv:hep-ph/0607020 [hep-ph]].

\bibitem{Antusch:2014woa}
  S.~Antusch and O.~Fischer,
  JHEP \textbf{10} (2014), 094
  [arXiv:1407.6607 [hep-ph]].

\bibitem{Calibbi:2017uvl}
L.~Calibbi and G.~Signorelli,
Riv. Nuovo Cim. \textbf{41} (2018) no.2, 71-174
[arXiv:1709.00294 [hep-ph]].

\bibitem{Jarlskog:1985ht}
C.~Jarlskog,
Phys. Rev. Lett. \textbf{55} (1985), 1039.

\bibitem{Cheng:1980tp}
T.~P.~Cheng and L.~F.~Li,
Phys. Rev. Lett. \textbf{45} (1980), 1908.

\bibitem{Inami:1980fz}
T.~Inami and C.~S.~Lim,
Prog. Theor. Phys. \textbf{65} (1981), 297.

\bibitem{Lim:1981kv}
C.~S.~Lim and T.~Inami,
Prog. Theor. Phys. \textbf{67} (1982), 1569.

\bibitem{Langacker:1988up}
P.~Langacker and D.~London,
Phys. Rev. D \textbf{38} (1988), 907.

\bibitem{Fernandez-Martinez:2016lgt}
E.~Fernandez-Martinez, J.~Hernandez-Garcia and J.~Lopez-Pavon,
JHEP \textbf{08} (2016), 033
[arXiv:1605.08774 [hep-ph]].

\bibitem{Goswami:2008mi}
S.~Goswami and T.~Ota,
Phys. Rev. D \textbf{78} (2008), 033012
[arXiv:0802.1434 [hep-ph]].

\bibitem{Drewes:2013gca}
M.~Drewes,
Int. J. Mod. Phys. E \textbf{22} (2013), 1330019
[arXiv:1303.6912 [hep-ph]].

\bibitem{Hisano:1995cp}
J.~Hisano, T.~Moroi, K.~Tobe and M.~Yamaguchi,
Phys.Rev.D\textbf {53} (1996), 2442 - 2459 [arXiv:hep-ph/9510309[hep-ph]].

\bibitem{Hisano:1995nq}
J.~Hisano, T.~Moroi, K.~Tobe, M.~Yamaguchi and T.~Yanagida,
Phys.Lett.B\textbf {357} (1995), 579 - 587 [arXiv:hep-ph/9501407[hep-ph]].

\bibitem{Xing:2020ald}
Z.~z.~Xing and Z.~h.~Zhao,
[arXiv:2008.12090 [hep-ph]].

\bibitem{Buras:1989xd}
A.~J.~Buras and P.~H.~Weisz,
Nucl. Phys. B \textbf{333} (1990), 66-99.

\bibitem{tHooft:1972tcz}
G.~'t Hooft and M.~J.~G.~Veltman,
Nucl. Phys. B \textbf{44} (1972), 189-213.

\bibitem{Breitenlohner:1977hr}
P.~Breitenlohner and D.~Maison,
Commun. Math. Phys. \textbf{52} (1977), 11-38.

\bibitem{Breitenlohner:1975hg}
P.~Breitenlohner and D.~Maison,
Commun. Math. Phys. \textbf{52} (1977), 39.

\bibitem{Breitenlohner:1976te}
P.~Breitenlohner and D.~Maison,
Commun. Math. Phys. \textbf{52} (1977), 55.

\bibitem{tHooft:1978jhc}
  G.~'t Hooft and M.~J.~G.~Veltman,
  Nucl. Phys. B \textbf{153} (1979), 365-401.

\bibitem{Passarino:1978jh}
  G.~Passarino and M.~J.~G.~Veltman,
  Nucl. Phys. B \textbf{160} (1979), 151-207.

\bibitem{Denner:1991kt}
  A.~Denner,
  Fortsch. Phys. \textbf{41} (1993), 307-420
  [arXiv:0709.1075 [hep-ph]].

\bibitem{Patel:2015tea}
  H.~H.~Patel,
  Comput. Phys. Commun. \textbf{197} (2015), 276-290
  [arXiv:1503.01469 [hep-ph]].

\bibitem{Patel:2016fam}
  H.~H.~Patel,
  Comput. Phys. Commun. \textbf{218} (2017), 66-70
  [arXiv:1612.00009 [hep-ph]].

\end{thebibliography}
\end{document}